\documentclass[aps,twocolumn,prd,reprint,superscriptaddress,nofootinbib,floatfix]{revtex4-2}
\RequirePackage[l2tabu, orthodox]{nag}

\usepackage[utf8x]{inputenc}
\usepackage{microtype}

\usepackage{amsmath}
\usepackage{amstext, amssymb, amsthm, amsfonts, slashed}
\usepackage{mathtools}
\usepackage[version=4]{mhchem}
\usepackage{graphicx}
\usepackage{booktabs}
\usepackage{dcolumn}
\usepackage{bm}
\usepackage{dsfont}
\usepackage[usenames,dvipsnames]{xcolor}
\usepackage[normalem]{ulem}

\usepackage{url}
\usepackage[hypertexnames]{hyperref}
\usepackage[nameinlink]{cleveref}
\crefname{figure}{Fig.}{Figs.}
\Crefname{figure}{Figure}{Figures}
\crefname{section}{Sec.}{Secs.}
\Crefname{section}{Section}{Sections}
\crefname{equation}{Eq.}{Eqs.}
\Crefname{equation}{Equation}{Equations}
\crefname{table}{Table}{Tables}
\Crefname{table}{Table}{Tables}
\crefname{appendix}{Appendix}{Appendices}
\crefname{subappendix}{Appendix}{Appendices}
\Crefname{appendix}{Appendix}{Appendices}
\Crefname{subappendix}{Appendix}{Appendices}
\usepackage{braket}
\usepackage{simplewick}
\usepackage{float}
\usepackage{subcaption}
\usepackage{physics}

\hypersetup{
	colorlinks,
	linkcolor={red!75!black},
	citecolor={blue!75!black},
	urlcolor={blue!75!black},
}

\renewcommand{\tr}{{\textrm{tr}}}

\newcommand{\orcid}[1]{\href{https://orcid.org/#1}{\includegraphics[height=1.7ex,width=1.7ex]{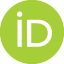}}}

\begin{document}

\title{Hadronic resonances from the 4PI effective action of QCD}

\author{Jonas~Wessely~\orcid{0009-0006-8700-104X}}
\thanks{\href{mailto:jonas.wessely@uni-giessen.de}{jonas.wessely@uni-giessen.de}} 
\affiliation{Institut für Theoretische Physik, Justus-Liebig–Universität Giessen, Heinrich-Buff-Ring 16, 35392 Giessen, Germany}
\author{Markus Q. Huber \,\orcid{0000-0002-7931-9801}}
\thanks{\href{mailto:markus.huber@theo.physik.uni-giessen.de}{markus.huber@theo.physik.uni-giessen.de}} 
\affiliation{Institut für Theoretische Physik, Justus-Liebig–Universität Giessen, Heinrich-Buff-Ring 16, 35392 Giessen, Germany}

\begin{abstract}
	We discuss the 4PI effective action as a tool to study resonances in QCD.
    To this end, we present the 4PI effective action of QCD in linear covariant gauges up to four loops and derive from it the kernels for two-body bound states, using mesons as an example. The emergence of resonances via decay channels is built into the formalism at the level of the four-point functions, and we employ an on-shell decomposition of its physical channels to arrive at a diagrammatic representation for physical two-body decays within the Bethe-Salpeter equation framework. We apply this formalism to the $\rho$, $K^*$ and $\phi$ mesons and extract their widths.
\end{abstract}

\maketitle

\section{Introduction}
\label{sec:intro}
The spectrum of quantum chromodynamics (QCD) consists of hadronic bound states and resonances, while particle states of the fundamental fields are barred from the physical spectrum due to confinement. The only hadrons composed of light quarks that are stable under the strong interaction are the pions and kaons together with the lightest baryons. All other hadrons are resonances with finite lifetimes and decay via the strong interaction. 

The theoretical description of resonances from first principles is challenging. It requires both non-perturbative methods and a framework to compute time-like correlations. A fairly successful approach has been the combination of lattice QCD and low-energy effective theories, two methods each of which covers one of these aspects. For reviews, see, e.g., \cite{Briceno:2017max,Hansen:2019nir,Mai:2022eur}. Functional methods, such as Dyson-Schwinger equations (DSEs), the functional renormalisation group (fRG), and $n$-particle irreducible ($n$PI) effective actions, on the other hand, are a natural choice for this particular combination of requirements, see, for instance, \cite{Alkofer:2000wg,Berges:2004pu,Fischer:2006ub, Pawlowski:2005xe, Gies:2006wv, Carrington:2010qq,Swanson:2010pw,Dupuis:2020fhh,Eichmann:2025wgs,Huber:2025cbd}. In contrast to lattice QCD, they are formulated in the continuum and are in principle not limited to Euclidean spacetime. In practice, the resolution of the analytic structure of correlation functions in the complex plane remains challenging, but substantial progress has been made in recent decades. For an incomplete list, see~\cite{Horak:2020eng, Horak:2021pfr, Fehre:2021eob, Horak:2022myj, Horak:2022aza, Braun:2022mgx, Horak:2023hkp, Eichmann:2023tjk, Pawlowski:2024kxc, Kockler:2025kdt, Pawlowski:2025etp, Kockler:2026zkj} for (generalised) spectral representations, \cite{Floerchinger:2011sc, Kamikado:2013sia, Tripolt:2013jra, Pawlowski:2015mia, Jung:2016yxl, Pawlowski:2017gxj, Tripolt:2018jre, Tripolt:2018qvi, Corell:2019jxh, Heller:2021wan} for analytically continued and temporal fRG flows, \cite{Gasenzer:2007za, Gasenzer:2010rq, Huelsmann:2020xcy, Tan:2021zid, Roth:2021nrd, Roth:2023wbp,Roth:2024rbi,Noel:2026clb} for computations on the Schwinger-Keldysh contour, and \cite{Weil:2017knt, Williams:2018adr, Eichmann:2019dts, Miramontes:2019mco, Santowsky:2020pwd, Eichmann:2021vnj, Miramontes:2021xgn, Santowsky:2021ugd, Fischer:2020xnb, Huber:2022nzs} for contour deformations.

For the functional derivation of scattering kernels and bound state equations, the $n$PI formalism provides a natural framework, and the 3PI effective action has been applied successfully to the computation of glueballs in Yang-Mills theory \cite{Huber:2021yfy} and of mesons \cite{Williams:2015cvx}. The extension to 4PI is especially important for the description of resonances, as physical two-body states into which a particle can decay appear as dynamically generated poles of four-point functions.

In this article, we present a general framework to compute hadronic decay widths in functional QCD.
We use two-body decays of vector mesons as an example, but the framework is not limited to them. 
It is built in three major steps: first, we present the 4PI effective action of QCD in linear covariant gauges up to four loops and derive the two-body scattering kernels from it. The second step is to decompose the four-point functions into their physical channels, where we employ an on-shell approximation that replaces the four-point functions by their on-shell poles and residues. This allows us to identify the diagrams that contain the physical thresholds of the decay products. In the last step, we set up a computational framework to solve the homogeneous Bethe-Salpeter equation (BSE) for resonances by means of contour deformations of the integration path in the complex plane. We provide a general algorithm to find an optimised contour and present results for the $\rho$, $K^*$ and $\phi$ mesons.

The paper is structured as follows: we briefly review the $n$PI formalism and present the 4PI effective action of QCD in \cref{sec:nPIFormalism}. There, we also discuss the equations of motion (EOMs) of correlation functions and the two-quark scattering kernel. This is followed by a discussion of the emergence of resonances and decay channels in \cref{sec:Resoncances}, where we define and discuss the on-shell decomposition to arrive at the decay diagrams in the BSE. Special emphasis is put on the mixing of states. The application of the formalism to light vector mesons is presented in \cref{sec:VMDecays}, where we discuss the necessary ingredients and the numerical computation of the decay kernels. Our results for the $\rho$, $K^*$ and $\phi$ mesons are presented in \cref{sec:results}. There, we also draw the connection to triangle diagrams and other results in the literature. We conclude in \cref{sec:conclusion} with a summary. Several technical details are deferred to the appendices.

\section{The $n$PI formalism}
\label{sec:nPIFormalism}

We revisit the $n$PI effective actions and the corresponding EOMs and scattering kernels and present the 4PI effective action of QCD in linear covariant gauges up to four loops.

\subsection{$n$PI effective actions}
\label{sec:nPI}

We provide a brief review of the $n$PI formalism (${n>1}$). Starting from a path integral representation of the generating functional of connected correlation functions
\begin{align}\label{eq:def_W}
	W[J] = \ln \int \mathcal{D}\phi \, e^{-S[\phi] + J^i  \phi_i}\,,
\end{align}
one can perform a Legendre transformation to obtain the 1PI effective action:
\begin{align}\label{eq:def_1PI}
	\Gamma_{1\text{PI}}[\Phi] = -W[J] + J^i \frac{\delta W[J]}{\delta J^i}\,.
\end{align}
The field $\phi$ (and similarly $\Phi$) represents a superfield vector that contains all fundamental fields of the theory and $J$ is the vector of the corresponding sources. We use a shorthand notation where a single index represents the field type, the spacetime coordinate or momentum, and all discrete indices.
Repeated indices mean summation and integration.
The mean field $\Phi$ is the expectation value of the superfield $\phi$ in the presence of the source $J$,
\begin{align}
\Phi=\langle \phi\rangle_J=\frac{\delta W[J]}{\delta J}.
\end{align}
The EOMs of the 1PI effective action are the DSEs, which can be obtained as the Ward identities of the shift symmetry of the path integral measure.
An extension of the 1PI effective action, the average effective action, is the central object in the functional renormalisation group (fRG). For reviews and pedagogical introductions, see \cite{Alkofer:2000wg, Fischer:2006ub, Pawlowski:2005xe, Gies:2006wv, Swanson:2010pw,Dupuis:2020fhh,Huber:2025cbd}.
Both methods require only the classical action as input to generate a full set of equations for the 1PI correlation functions, which are the moments of the 1PI effective action. These integral or integro-differential equations are built from fully dressed correlation functions.
While the fRG is one-loop exact, DSEs contain up to two loops in QCD. However, the equations do not close at any finite order of correlation functions but form an infinite tower. Hence, truncations are necessary to obtain a finite set of equations. 

A complementary approach is the $n$PI formalism, in which additional source terms for propagators and higher $n$-point functions are introduced \cite{Cornwall:1974vz,Carrington:2004sn,Berges:2004pu,Carrington:2010qq}: 
\begin{align}\label{eq:def_W_nPI}
	W[J,R_2,R_3,\dots] = \ln \int \mathcal{D}\phi \, e^{-S[\phi] + J^i \phi_i + \frac{1}{2!}R_2^{ij} \phi_i \phi_j + \dots}\,.
\end{align}
Here, $R_2^{ij}$ are sources for the propagators, $R_3^{ijk}$ for the three-point functions, and so on.
Note that it is not necessary to include source terms for all operators at a given order, as the schematic notation in \labelcref{eq:def_W_nPI} might suggest. In particular, one might not include source terms for operators that are forbidden by symmetries of the theory, such as a three-point function of quarks in QCD. However, such sources can be included for the derivation of corresponding scattering kernels. Moreover, the operators that are coupled to a source can carry non-trivial quantum numbers encoded in their tensor structures. 

As in the 1PI case, one obtains the $n$PI effective action by performing successive Legendre transformations with respect to additional sources. Schematically, this reads
\begin{align}\label{eq:def_Gamma_nPI}\nonumber
	&\Gamma_{n\text{PI}}[\Phi,G,V_3, \ldots ] =- W[J,R_2,\ldots] + J^i \frac{\delta W[J,R_2,\ldots]}{\delta J^i} \\[2ex]
	&+ R_2^{ij}  \frac{\delta W[J,R_2,\ldots]}{\delta R_2^{ij}} + R_3^{ijk}  \frac{\delta W[J,R_2,\ldots]}{\delta R_3^{ijk}} + \dots\,,
\end{align}
where $G$ and $V_n$ are the connected propagators and proper $n$-point vertices, respectively.
Typically, a systematic expansion in loops or $1/N$ \cite{Berges:2001fi} is employed. The EOMs for the respective $n$-point functions are given by the stationarity conditions of the $n$PI effective action which, at vanishing sources indicated by the subscript $_0$, read
\begin{align}\label{eq:eom_nPI}\nonumber
	\frac{\delta \Gamma_{n\text{PI}}[\Phi,G,V_3, \ldots]}{\delta G}\Bigg|_0 = 0\,,\\[2ex]\nonumber
	\frac{\delta \Gamma_{n\text{PI}}[\Phi,G,V_3, \ldots]}{\delta V_3}\Bigg|_0 = 0\,,\\\nonumber
	\vdots \notag \\
	\frac{\delta \Gamma_{n\text{PI}}[\Phi,G,V_3, \ldots]}{\delta V_n}\Bigg|_0 = 0\,.	
\end{align}
Given an explicit expression for the $n$PI effective action, \labelcref{eq:eom_nPI} provides a closed set of equations for all sourced $n$-point functions.
Non-sourced $n$-point functions do not appear, but their effects are encoded in loop diagrams.
In contrast to the DSEs, this is a finite set of equations, because the truncation is realised at the level of the action.

\subsection{The 4PI effective action in QCD}
\label{sec:4PI_QCD}

\begin{figure*}[ht]
	\centering
	\includegraphics[width=0.79\textwidth]{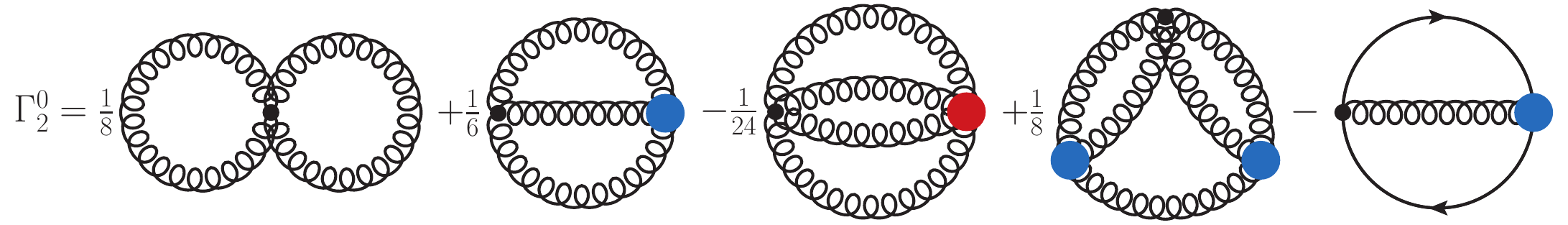}
		\caption{Diagrammatic representation of $\Gamma_2^0$, the part of the effective action that contains bare vertices. Bare vertices are denoted by small black dots, full three-point functions by larger blue dots, and full four-point functions by larger red dots. Wiggly lines are full variational gluon propagators and continuous lines fermionic (ghost/quark) propagators.\hspace*{\fill}}
		\label{fig:Gamma0}
\end{figure*}

\begin{figure*}[ht]
	\centering
	\includegraphics[width=0.92\textwidth]{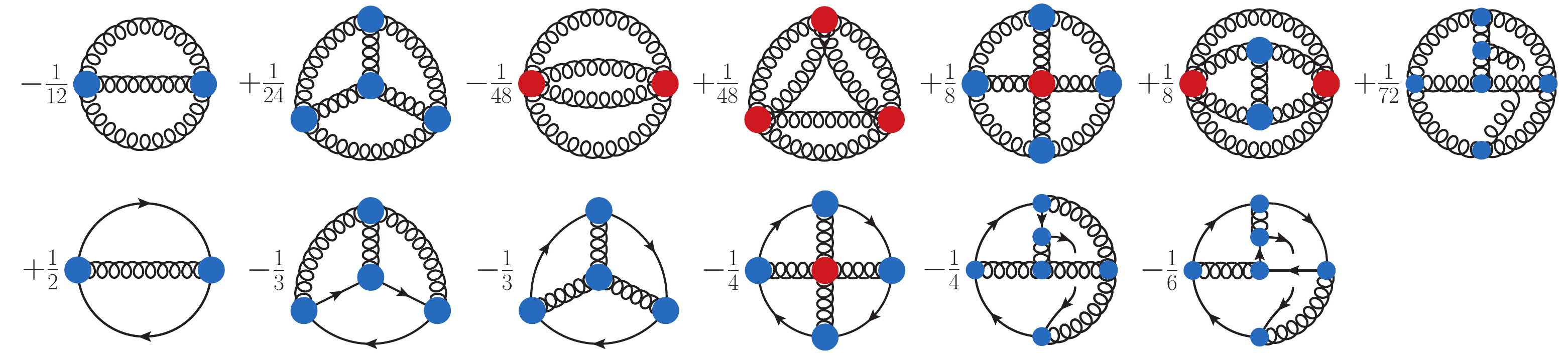}
		\caption{Diagrammatic representation of the part of $\Gamma_2^\text{int}$ that contains only primitively divergent vertices. Diagrams with fermionic loops represent the sum over all fermionic degrees of freedom (ghosts/quarks).\hspace*{\fill}}
		\label{fig:Gammaclverts}
\end{figure*}

\begin{figure*}[ht]
	\centering
	\includegraphics[width=0.90\textwidth]{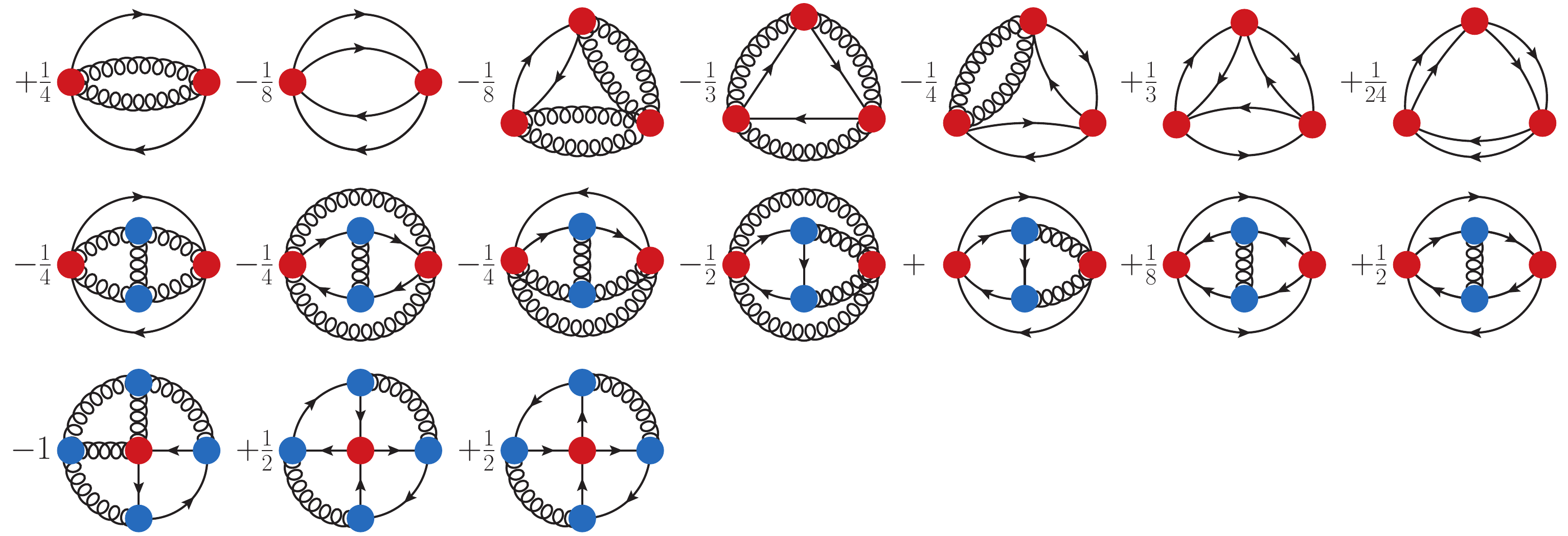}
		\caption{Diagrammatic representation of the part of $\Gamma_2^\text{int}$ that contains the two-fermion-two-gluon vertices and four-fermion vertices, viz., four-point functions that are not primitively divergent. \hspace*{\fill}}
		\label{fig:Gamma2g2fverts}
		\label{fig:gamma4fermipure}
\end{figure*}

\begin{figure}[t]
	\includegraphics[width=0.48\textwidth]{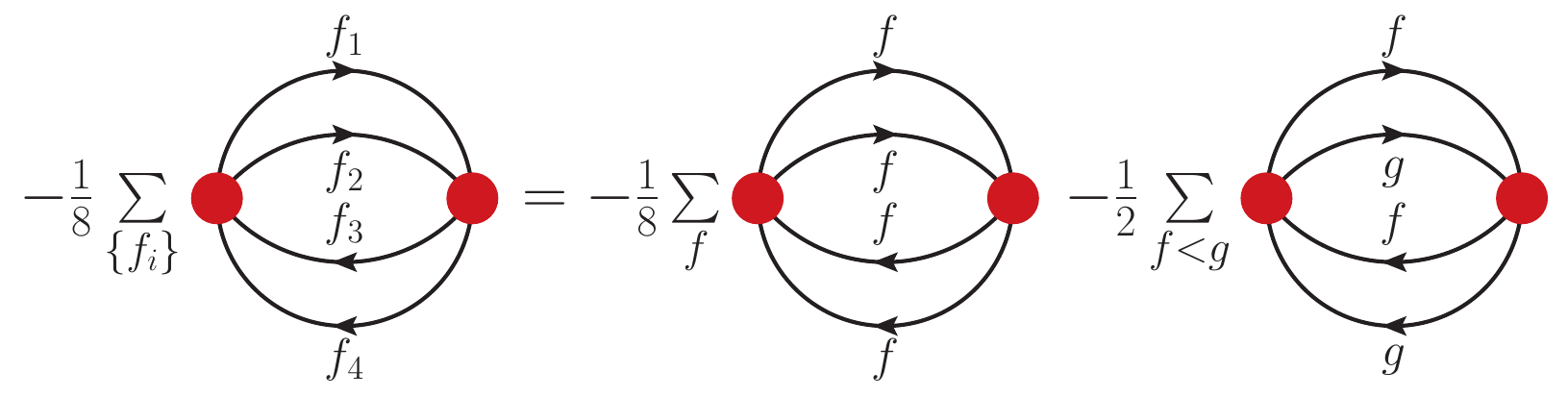}
		\caption{Summing the four-fermion basketball diagram over the flavours $f_i$ of all four fermion lines yields the single-flavour and the two-flavour diagrams with their respective coefficients.\hspace*{\fill}}
		\label{fig:flavoursum}
\end{figure}

Explicit expressions for the $n$PI effective action can be obtained by performing the Legendre transformations in \labelcref{eq:def_Gamma_nPI} order by order in a loop expansion. This has been done for scalar theories up to 5PI \cite{Carrington:2010qq} and for QCD up to 3PI \cite{Berges:2004pu}. In this section, we present the 4PI effective action of QCD in linear covariant gauges up to four loops. We start by discussing its derivation and present the resulting diagrams in \cref{fig:Gamma0,fig:Gamma2g2fverts,fig:Gammaclverts}.

Importantly, the 4PI topologies that were first identified in \cite{Carrington:2004sn} for a scalar theory directly carry over to other theories, including QCD, by invoking the superfield formalism. More precisely, the construction of the 4PI effective action in \cite{Carrington:2004sn} was already carried out in a condensed notation where the field indices are not explicitly written out, and its extension to superfields is straightforward. Hence, we take the topologies of the 4PI effective action of the scalar theory as a starting point from which an algorithmic construction of the 4PI effective action of QCD can be achieved by simply promoting the indices of the propagators and vertices to superindices that include the field type. Then, a summation over all fields for each internal line yields the four-loop 4PI diagrams of QCD after eliminating the vertices that are forbidden by symmetries. The summation and identification of identical diagrams also yields the correct symmetry factors of the diagrams. Alternatively, one can write down all non-identical diagrams for a given topology. Then, the symmetry factors can be obtained by computing the order of the automorphism group of each diagram. Minus signs are added for closed fermion loops. Here, one has to be careful with diagrams containing fermionic four-point functions, as their signs are not uniquely determined by a purely graphical representation. We fixed this by comparing the EOMs with the respective DSEs derived in \textit{DoFun} \cite{Alkofer:2008nt,Huber:2011qr,Huber:2019dkb}.

The 4PI effective action can be decomposed into a classical part $S_{\text{cl}}$, which is the classical gauge-fixed action evaluated on the mean field $\Phi$, the one-loop part $\Gamma_1$, which carries only the propagators and their classical inverses, and the interaction part $\Gamma_2$, which contains all diagrams with at least one vertex, see also \cite{Cornwall:1974vz}: 
\begin{align}\label{eq:4PI_decomposition}
	&\!\!\!\Gamma[\Phi,G,V_3,V_4] = S_{\text{cl}}[\Phi] + \Gamma_1[\Phi,G] - \Gamma_2[\Phi,G,V_3,V_4]\,,\\[2ex]\nonumber
	&\!\!\!\Gamma_2[\Phi,G,V_3,V_4] = \Gamma_2^0[\Phi,G,V_3,V_4] + \Gamma_2^{\text{int}}[G,V_3,V_4]	
	\,.
\end{align}
Here and in the following, we drop the subscript $4\text{PI}$ for brevity.
The one-loop part reads 
\begin{align}\label{eq:4PI_one_loop}
	\Gamma_1[\Phi,G] = &\frac{1}{2}\Tr\ln G_B^{-1} + \frac{1}{2}\Tr G_{B,0}^{-1}G_B \\[2ex]\nonumber
	& - \Tr\ln G_F^{-1} - \Tr G_{F,0}^{-1}G_F\,,
\end{align}
where $G_{B,0}^{-1}$ and $G_{F,0}^{-1}$ are the classical inverse propagators of bosons and fermions, respectively, and the trace is understood to be the supertrace over all (bosonic/fermionic) fields and indices.
We use the term \textit{fermionic} for all anticommuting fields, although the ghost field is a scalar.
Note that we separated the bosonic and fermionic contributions to the one-loop part to account for the relative minus signs. 
In components, $G_B$ contains the gluon propagator $G^{AA}_{\mu\nu}$ and $G_F$ the ghost and quark propagators $G^{c\bar c}$ and $G^{q\bar q}$, respectively. The vertices $V_3$ and $V_4$ represent the various three- and four-point functions. Individual vertices are labelled by their field content, e.g.,\ $V^{\bar q q A}_\mu$ for the quark-gluon vertex or $V^{\bar q q \bar q q}$ for a four-quark vertex. 

The interaction part $\Gamma_2$ can be further decomposed into contributions that contain bare vertices, $\Gamma_2^0$, and contributions that contain only full vertices, $\Gamma_2^{\text{int}}$.
The diagrams of $\Gamma_2^0$ in \cref{fig:Gamma0} are identical to those of the 3PI effective action except for a dressed four-gluon vertex and a different prefactor in the third diagram.
To keep the diagrams tractable, we group those of $\Gamma_2^{\text{int}}$ into two classes. The first class of diagrams, shown in \cref{fig:Gammaclverts}, contains only diagrams with vertices that have a classical counterpart.
Hence, they are also contained in the 2PI or 3PI effective action expanded to four loops, up to the dressing of the vertices and different symmetry factors. 
The second class also contains four-point functions with no bare counterpart, viz., two-fermion-two-gluon vertices and four-fermion vertices. In particular, the first two diagrams play a pivotal role in the derivation of the two-body kernels that facilitate the resolution of decays in the BSE as discussed in \cref{sec:decaydiags}. 

For fermion loops, it is important to remember that they represent the sum over all fermion flavours, including the ghosts. This summation also leads to the correct symmetry factors of the respective diagrams. We make this explicit in \cref{fig:flavoursum} with the example of the second diagram from \cref{fig:Gamma2g2fverts}, the fermion basketball. We can split this diagram into a single-flavour contribution without mixed four-fermion vertices and mixed basketballs. For the latter, we can find four different but equivalent combinations of fermion loops, leading to the correct symmetry factor of $1/2$. 

%
\subsection{Equations of motion and scattering kernels}
\label{sec:eoms}
We discuss the derivation of the EOMs for the correlation functions and the scattering kernels from the 4PI effective action. We recall that the resulting EOMs for the inverse propagators agree with the respective DSEs due to cancellations when enforcing the EOMs of the vertices.
The scattering kernels from the 4PI effective action will be of particular interest for the computation of bound states and resonances. The diagrammatic derivations, in particular the functional derivatives and the replacement of vertices by their EOMs, were carried out with \textit{DoFun}.

%
\subsubsection{Three- and four-point functions}
\label{sec:vertex-eoms}

We start with the vertex EOMs because they are needed in the propagator EOMs.
To derive the equations for the three- and four-point functions, we take the respective functional derivatives of \labelcref{eq:4PI_decomposition} and set them to zero. Since neither the classical action nor $\Gamma_1$ depends on $V_3$ or $V_4$, the EOMs \labelcref{eq:eom_nPI} for the higher-order vertices reduce to 
\begin{align}
	\frac{\delta \Gamma_2[\Phi,G,V_3,V_4]}{\delta V_3} = 0\,, \qquad
	\frac{\delta \Gamma_2[\Phi,G,V_3,V_4]}{\delta V_4} = 0\,.
\end{align}
A crucial role is played by the topologies that connect two $n$-point vertices via $n$ propagators, producing the sunset for three-point functions and the basketball for four-point functions. Differentiating these diagrams with respect to a (dressed) vertex produces the proper $n$-point function from the diagram in which both vertices are dressed (\cref{fig:Gammaclverts}), and the bare vertex from the diagram with one dressed vertex (\cref{fig:Gamma0}). For the four-point function, all diagrams that contribute beyond the bare vertex are of four-loop order in the effective action. Hence, the 4PI effective action truncated at three loops is equivalent to the 3PI effective action at three loops. This is an example of the equivalence hierarchy in $n$PI effective actions \cite{Berges:2004pu}.
%
\subsubsection{Two-point functions}
\label{sec:eoms-2pt}
The EOMs for the inverse propagators $G^{-1}$ are obtained by taking the functional derivative of~\labelcref{eq:4PI_decomposition} with respect to $G$ and setting the sources to zero. Distinguishing between bosonic and fermionic fields, this yields 
\begin{subequations}
\label{eq:eom_2pt}
\begin{align}
	G_B^{-1} &= G_{B,0}^{-1} - 2 \frac{\delta \Gamma_2[\Phi,G,V_3,V_4]}{\delta G_B}\,, \\
	G_F^{-1} &= G_{F,0}^{-1} + \frac{\delta \Gamma_2[\Phi,G,V_3,V_4]}{\delta G_F}\,.
\end{align}
\end{subequations}
The signs and factors for bosons and fermions are trivially read off from the one-loop part in \labelcref{eq:4PI_one_loop}.
Using the EOMs for the vertices after performing the derivatives, one can show that Equations (\ref{eq:eom_2pt}) agree with the respective DSEs. For the 4PI effective action of a scalar theory, this is discussed in \cite{Carrington:2004sn,Carrington:2010qq}, but the structural analysis directly carries over and can be shown algorithmically. Details, including the implementation in \textit{DoFun} \cite{Alkofer:2008nt,Huber:2011qr,Huber:2019dkb}, will be provided elsewhere.

%
\subsubsection{Scattering kernels}
\label{sec:scattering-kernels}

\begin{figure}[t]
	\includegraphics[width=0.48\textwidth]{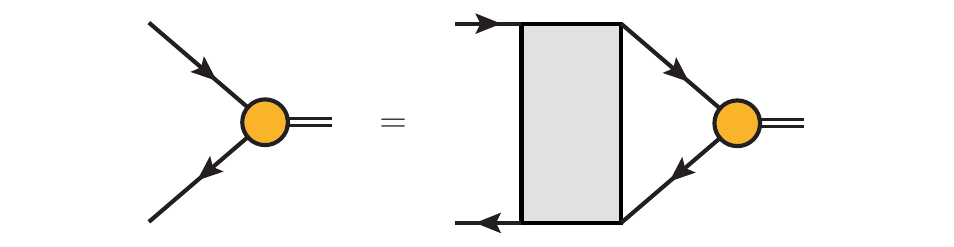}
		\caption{Homogeneous BSE for fermionic constituents. The lines connecting to the amplitude on the right-hand side are superfield propagators, allowing for mixing. \hspace*{\fill}}
		\label{fig:homogenous_BSE}
\end{figure}

The spectrum of a theory is encoded in the T-matrix, which is the non-trivial part of the S-matrix. It fulfils the scattering equation
\begin{align}\label{eq:scattering_equation}
	T = K + K G_0 T\,,
\end{align}
where $K$ is the scattering kernel and $G_0$ the product of the propagators of the fields involved in the scattering process. This is valid for bound states with an arbitrary number of constituents. 
Close to a resonance or a bound state, the T-matrix is dominated by the corresponding pole at $P^2 = -m^2$. Schematically, this reads for a two-body bound state
\begin{align}
	T_{ij;kl} \to \frac{\Gamma_{ij} \overline\Gamma_{kl} }{P^2 + m^2}\,,
\end{align}
with the Bethe-Salpeter amplitudes $\Gamma_{ij}$ and $\overline\Gamma_{kl}$, where the bar indicates charge conjugation. Then, the scattering equation \labelcref{eq:scattering_equation} reduces to a homogeneous BSE for the bound state amplitude
\begin{align}\label{eq:homogeneous_BSE}
	\Gamma_{ij} = K_{ij;kl} G_{kk'} G_{l'l} \Gamma_{k'l'}\,.
\end{align}
For a diagrammatic representation of the BSE with quark legs, see \cref{fig:homogenous_BSE}.

The scattering kernel fits naturally into the $n$PI framework, as it is obtained by taking two consecutive functional derivatives of the effective action with respect to the full $m$-point function of the constituents \cite{Fukuda:1987su,Komachiya:1989kc,McKay:1989rk}, see also \cite{Sanchis-Alepuz:2015tha} for a pedagogical introduction. 

For a two-body bound state, we perform two derivatives with respect to two-point functions.
We start with the first derivative, which is given in Equations~(\ref{eq:eom_2pt}).
The left-hand side yields two inverse propagators. In the final expression for the Bethe-Salpeter amplitude, they are absorbed into the Bethe-Salpeter wave function.
The right-hand side yields the scattering kernel
\begin{align}\label{eq:K}
		K_{ij;kl} = m_{b,f} \, \frac{\delta^2 \, \Gamma_2[\Phi,G,V_3,V_4]}{\delta G^{ij} \delta G^{kl}}\,\Bigg|_0,
\end{align}
where $m_{b,f}$ depends on whether the first derivative, with indices $i$ and $j$, acts on a bosonic ($m_b = 2$) or a fermionic ($m_f=-1$) propagator. The effect of a functional derivative with respect to a propagator in $\Gamma_2$ is to open an internal line of a vacuum diagram. Since the diagrams of $\Gamma_2$ are two-particle irreducible, the result stays connected, and an $L$-loop diagram of the effective action yields kernel diagrams with $L-2$ loops. Crucially, the second derivatives (with respect to the indices $k$ and $l$) need to be performed for all possible two-point functions, as they connect to the corresponding Bethe-Salpeter amplitudes. This allows mixing between different Bethe-Salpeter amplitudes, see \cite{Huber:2025kwy} for an example.
The possible coupling between different amplitudes is realised here via the summation over the intermediate indices in \labelcref{eq:K}, which includes the summation over different fields.
The sources are set to zero after the derivatives have been performed, so that only physical propagators remain.

%
\begin{figure*}[t]
	\includegraphics[width=0.8\textwidth]{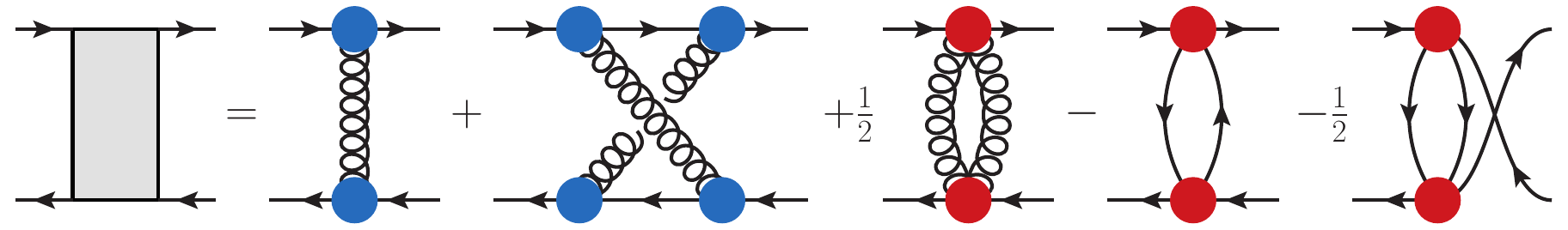}
		\caption{Meson kernel truncated at one-loop order. Flavour-changing interactions are implied by the summation over the quark lines connecting to the meson Bethe-Salpeter amplitude. 
		 \hspace*{\fill}}
		\label{fig:meson-kernel}
\end{figure*}

\section{Resonances in the 4PI framework}
\label{sec:Resoncances}
In this section, we discuss the emergence of resonances and bound states in the 4PI framework. We start by introducing them as poles in the elementary four-point functions and discuss the mixing of states.
In a second step, we shift the focus to the determination of resonance masses and, in particular, decay widths with homogeneous BSEs. There, we use the 4PI scattering kernels to derive the diagrammatic decay channels in the BSE. 

\subsection{Decay products as on-shell states}

\begin{figure}[t]
	\includegraphics[width=0.48\textwidth]{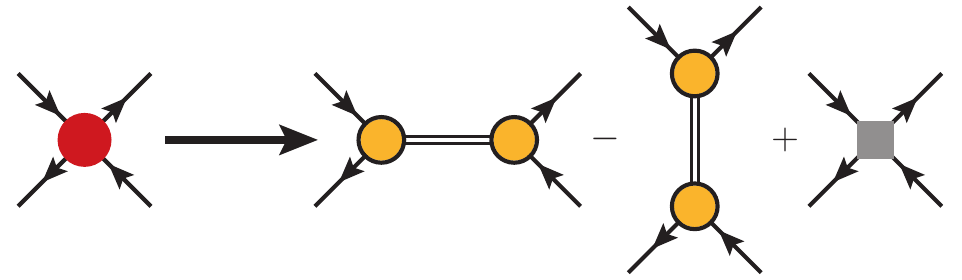}
		\caption{Separation of the quark four-point function (red~disk) into mesonic on-shell poles, indicated by double lines, and the rest, indicated by a grey box. The yellow disks represent the Yukawa couplings of the mesons to the quarks, defined by the respective Bethe-Salpeter amplitudes.\hspace*{\fill}}
		\label{fig:onshell-replacements}
\end{figure}

In the functional formalism, bound states can be described by $n$-body bound state equations, the simplest realisation of which are the two-body BSEs. This applies to true bound states as well as to resonances. In this work, we exclusively discuss two-body equations. They are based on the fact that two-body bound states appear as poles in the four-point functions of the fundamental constituents.
To describe the decays of resonances, diagrams involving the corresponding $n$-point functions must be included in the scattering kernel of a BSE. For example, to describe the decay of a $\rho$ meson into two pions, we need to include diagrams with at least two quark-antiquark four-point functions that contain the corresponding pion poles. The first and dominant class of such diagrams appears at one-loop order in the two-quark scattering kernel, see diagrams 4 and 5 in \cref{fig:meson-kernel}. At this point, we emphasise that two-particle states can be related to gauge-variant operators, as in the case of quark-antiquark states, but also to mere components of physical operators. An example of the latter is the simplest glueball operator, $F^2$, which contains two-gluon, three-gluon and four-gluon components. However, we consider the existence of a pole in one of them as a sufficient criterion for a pole of the full operator, as it could only be removed via intricate cancellations with the other components \cite{Pawlowski:2005xe,Huber:2021yfy}.

In QCD and Yang-Mills theory, the decay products need to be physical states, as the decay channels that are potentially opened by decays into the elementary quark and gluon fields are ultimately forbidden by confinement. The question of how the branch cuts arising from such unphysical channels are indeed cancelled is still open. A possible resolution is provided by a non-perturbative BRST quartet mechanism \cite{Kugo:1979gm,Alkofer:2011pe,Comitini:2026xrq}. The 4PI framework is a convenient starting point for incorporating such quartet states, as we discuss at the end of \cref{sec:decaydiags}. 

The decay thresholds are produced by the on-shell poles of the mesons in the four-quark vertex. Without approximation, we can write the latter as a sum of such on-shell poles and the rest (all momenta incoming):
\begin{align}\label{eq:4q-onshell-decomposition}\nonumber
	&V^{\bar q q \bar q q}(p_1,p_2,p_3,p_4)= \sum_{i} \Gamma^{\bar q q i}(p_1,p_2) D_i(s)\overline{\Gamma}^{\bar q q i}(p_3,p_4) \\[2ex]
&\hspace*{.5cm}- \text{t-channel} + \text{u-channel} + V^{\bar q q \bar q q}_\text{rest}(p_1,p_2,p_3,p_4)\,,
\end{align}
where $D_i(s)=\frac{1}{s+m_i^2}$ is the pole of a resonance in the s-channel with $s=(p_1+p_2)^2$ and $\overline{\Gamma}$ is the charge conjugate of the Bethe-Salpeter amplitude $\Gamma$. This decomposition should not be restricted to a single channel. By crossing symmetry, the same structure appears in the t-channel, defined here as $t=(p_1+p_4)^2$. It pairs the antiquark leg $1$ with the quark leg $4$ and the antiquark leg $3$ with the quark leg $2$, i.e., it follows from the s-channel term by exchanging the two quark legs, $p_2\leftrightarrow p_4$. Since the four-quark vertex is antisymmetric under the exchange of two fermion legs, the t-channel pole enters with a relative minus sign. These are the two channels in which a meson can form. The u-channel, on the other hand, here defined by $u=(p_1+p_3)^2$, carries diquark resonances, which, albeit not part of the physical spectrum, do contribute as off-shell degrees of freedom and become important in the description of baryons \cite{Eichmann:2016yit,Barabanov:2020jvn} or at finite baryon density \cite{Fukushima:2010bq}, see also \cite{Gholami:2026cgd}. The term indicated by the subscript ``rest'' contains all non-resonant interactions, for example, higher-order thresholds that are relevant for heavier resonances.

Our focus lies on mesonic intermediate states, in particular pions and kaons as the lightest stable (in QCD) bound states, but the extension to diquarks is straightforward. Keeping only the meson poles, we show a diagrammatic representation of \labelcref{eq:4q-onshell-decomposition} in \cref{fig:onshell-replacements}. The yellow disks represent the Yukawa couplings of the respective mesons to the quarks, formally defined by their Bethe-Salpeter amplitudes. The rest of the four-quark vertex is indicated by the grey box. 

\subsection{Mixing of states}\label{sec:mixingstates}
The mesons in the s- and t-channels are not necessarily the same. This depends on whether the four-fermion vertex has four identical legs or couples two different flavours. We illustrate this with two examples. For a four-strange vertex, both the s- and the t-channel contain a $\phi$ meson, which contains an $s \bar s $ pair. If, on the other hand, we consider a four-quark vertex with two up and two down quarks, $V^{\bar u u \bar d d}$, the lightest pole in the s-channel is that of the neutral pion, whose residue is given by the mixing of the up- and down-quark Bethe-Salpeter amplitudes, 
\begin{align}\nonumber 
	&V^{\bar u u \bar d d}(p_1,p_2,p_3,p_4)\big|_{s\to -m_{\pi^0}^2}\\[2ex]
	 &= \Gamma^{\bar u u \pi^0}(p_1,p_2) D_{\pi^0}(s)\overline{\Gamma}^{\bar d d \pi^0}(p_3,p_4)\,.
\end{align}
The t-channel, on the other hand, has a different flavour structure, and the lightest pole is that of the charged pion,
\begin{align}\nonumber
	&V^{\bar u u \bar d d}(p_1,p_2,p_3,p_4)\big|_{t\to -m_{\pi^\pm}^2} \\[2ex]
	&= - \Gamma^{\bar u d \pi^-}(p_1,p_4) D_{\pi^-}(t)\overline{\Gamma}^{\bar d u \pi^-}(p_3,p_2)\,.
\end{align}
This single term covers both charge orientations of the exchanged pion: a $\pi^-$ propagating from the pair $(p_1,p_4)$ to $(p_3,p_2)$ is the same exchange as a $\pi^+$ propagating in the opposite direction.
In the isospin-symmetric approximation, which we employ throughout this work, the two channels differ only in their flavour structure. This becomes important for the decay diagrams of the vector mesons, and we defer its discussion to \cref{app:DecayBSSymmetries,app:flavortrace}. 

This logic applies not only to mesonic poles in four-fermion vertices but also to other four-point functions. Generally speaking, the mixing of states at the level of four-point functions is realised by shared poles with identical quantum numbers, formalised by the tensor structures. What is not shared is the residue of the pole, which is given by the respective Bethe-Salpeter amplitudes. For example, the s-channel of a two-quark-two-gluon vertex contains the poles of isoscalar mesons, e.g., the $\eta'$, 
\begin{align}\nonumber
	&V^{\bar q q A A}(p_1,p_2,p_3,p_4)\big|_{s\to -m_{\eta'}^2} \\[2ex]
	&= \Gamma^{\bar q q \eta'}(p_1,p_2) D_{\eta'}(s)\overline{\Gamma}^{A A \eta'}(p_3,p_4)\,.
\end{align}
The residue $\overline{\Gamma}^{A A \eta'}$ is the gluonic component of the $\eta'$, which in this case is enhanced by the $U_A(1)$ anomaly, see, e.g.,~\cite{Cheng:2008ss,Alkofer:2008et}. If the mixing is weak, this amplitude is small and the contribution of the pole to the two-quark-two-gluon vertex is suppressed. For isovector states such as the $\pi^0$, it vanishes in the isospin-symmetric limit.
\begin{figure*}[t]
	\includegraphics[width=0.80\textwidth]{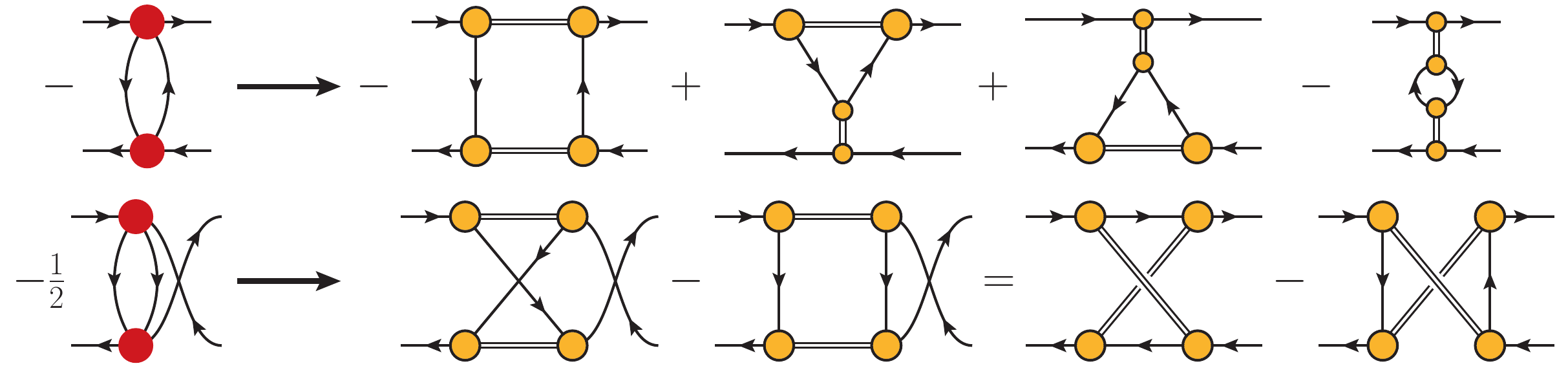}
		\caption{On-shell replacements in the kernel diagrams. The equation in the second row corresponds to a switch of the upper and lower vertices on the right-hand side. \hspace*{\fill}}
		\label{fig:onshellinsertion}
\end{figure*}

\subsection{Decay diagrams}\label{sec:decaydiags}

As discussed above, the meson kernel truncated at one-loop order, depicted in \cref{fig:meson-kernel}, carries the poles of physical states in the four-point functions and, if the latter are properly resolved, allows the resolution of decay widths of the two-body bound state. In practice, such a resolution of the four-point function in the time-like momentum region is not yet feasible, numerically or technically. Advances in this direction have been made on the Euclidean axis in QCD in \cite{Fu:2025hcm},
where the dependence on the s-, t- and u-channels of the four-quark vertex was included self-consistently. A similar analysis for complex momenta would require substantial technical advances. 

Instead, we utilise the on-shell decomposition \labelcref{eq:4q-onshell-decomposition}, where we approximate the four-point functions by their poles. The insertion of this decomposition into the fourth and fifth diagrams in \cref{fig:meson-kernel} is shown in \cref{fig:onshellinsertion}. The first row leads to the box diagram, where the two mesons are exchanged in the s-channel on the upper and lower line. The second and third diagrams are the cross terms, in which one meson is exchanged in the s-channel and the other in the t-channel. They do not contribute to the width and, together with similar diagrams that have a gluon exchange instead of the horizontal meson line, they can be resummed into a single vertical meson exchange. The last diagram in the first row shows two vertical t-channel exchanges connecting to a closed quark loop. This exchange is nothing but the second term of a Dyson series of the meson propagator, indicating that the inclusion of higher loop orders in the kernel, beyond those generated by the four-loop effective action, will lead to a kernel with a fully dressed meson exchange. While these diagrams are interesting from a conceptual point of view, they do not contribute to the decay width of the resonances we consider. For vector mesons, the impact of one-pion exchanges give was explored in \cite{Fischer:2008wy}. Here, we drop them in the following discussion. In summary, the two diagrams contributing to the width are the first one in the first row and the second diagram in the second row.

The second row of \cref{fig:onshellinsertion} leads to two types of crossed-box diagrams. In the first, two mesons are exchanged in a crossed ladder between the two quark lines, and in the second, two mesons are intermediate states in a u-channel type exchange. Note the relative minus sign, which is a direct consequence of the antisymmetry of the four-quark vertex under the exchange of two quark legs, i.e., the minus sign in front of the t-channel in \cref{fig:onshell-replacements}. Only the second diagram contributes to the two-body decay width of the resonance, while the first is a crossed-ladder pion exchange, a correction to the binding energy and expected to be subleading, given that even the one-pion exchange is only a minor correction.

We close this section with a remark on a potential realisation of confinement via BRST quartets. The quartet states are expected to be realised as certain bound states of the gauge-fixed constituents \cite{Kugo:1979gm,Alkofer:2011pe,Comitini:2026xrq}. In such a scenario, these states would enter the equations as intermediate states in the four-point functions, just like the physical states discussed above. For example, the two-ghost-two-gluon vertex contains poles in the s-channel that correspond to physical glueballs \cite{Huber:2020ngt}. Additionally, if the Schwinger mechanism is realised to generate the gluon mass gap \cite{Aguilar:2011xe, Aguilar:2015bud, Papavassiliou:2022wrb, Ferreira:2025anh}, it will also contain a pole in the s-channel associated with the corresponding massless excitation. Moreover, the t-channel of the two-ghost-two-gluon vertex could contain a pole that corresponds to the ghost-gluon bound state discussed in \cite{Alkofer:2011pe} as part of the non-perturbative BRST quartet. The inclusion of these states could cancel the unphysical cuts in observables that are introduced by diagrams with decays into elementary fields. 

\section{Vector meson decays}\label{sec:VMDecays}

In this section, we apply the framework developed above to the decays of vector mesons into light pseudoscalar mesons. We start with details on the DSE/BSE setup we use to solve the vector meson BSEs with the decay diagrams. This requires results of the quark propagator DSE for light and strange quarks and the Bethe-Salpeter amplitudes of the pseudoscalar mesons, all of them in the complex momentum plane. In \cref{sec:vector-mesons,sec:contour-deformations}, we proceed to the vector meson BSEs and discuss the structure of the decay diagrams and the contour deformations necessary to solve the BSEs above the decay thresholds. Special emphasis is put on the contours for decay products with different masses, as for the $K^*$ meson. 

\subsection{Setting up the ingredients}\label{sec:mesonBSEs}

\begin{figure*}[t]
	\begin{minipage}[t]{0.48\textwidth}
		\includegraphics[width=\textwidth]{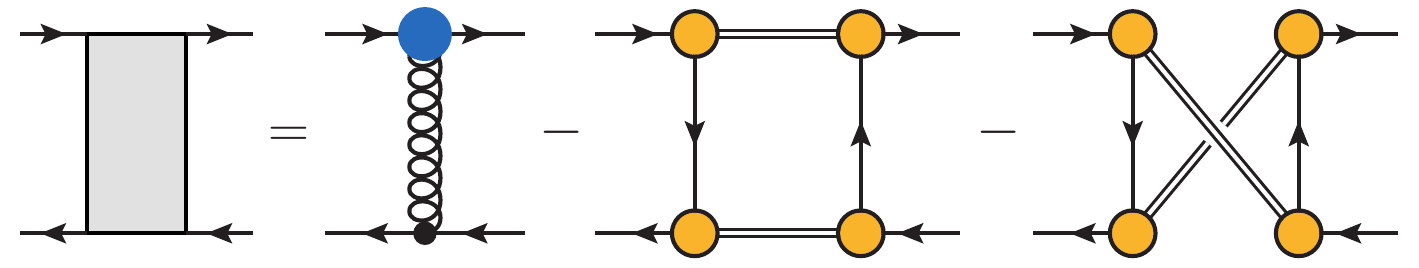}
		\caption{Truncated meson BSE kernel containing only the dominant one-gluon exchange and the derived two-body decay diagrams. The gluon exchange features only a single full vertex to be consistent with the gap equation in \cref{fig:quarkgap} in our truncation.\hspace*{\fill}}
		\label{fig:MesonBSE}
	\end{minipage}\hfill
	\begin{minipage}[t]{0.48\textwidth}
		\includegraphics[width=\textwidth]{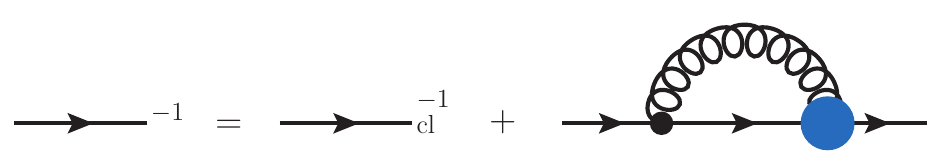}
		\caption{Quark gap equation. The subscript cl denotes the classical inverse quark propagator, the small black dot the bare quark-gluon vertex and the blue dot the full quark-gluon vertex; all internal propagators are full.\hspace*{\fill}}
		\label{fig:quarkgap}
	\end{minipage}
\end{figure*}

The homogeneous BSE and a two-quark scattering kernel were introduced in \cref{fig:homogenous_BSE,fig:meson-kernel}, respectively. The realisation of decay diagrams in the BSE was discussed in the previous section, resulting in the diagrams shown in \cref{fig:onshellinsertion}. On the diagrammatic level, we follow the truncation of \cite{Williams:2018adr} and only keep the gluon exchange and the meson boxes. The dropped vertical meson exchange was investigated in \cite{Fischer:2008wy} where, using certain approximations, its contribution to the binding energy was found to be roughly 100\,MeV. The crossed gluon exchange, on the other hand, which was found to be subleading, e.g., in \cite{Ferreira:2025wpu}, is suppressed by the gapping of the gluon propagators and by $1/N_c$ compared to a ladder two-gluon exchange. For completeness, we mention that we do not consider two-quark-two-gluon vertices: while the diagrams in \cref{sec:4PI_QCD} are kept general, we proceed without the corresponding sources. The resulting BSE is depicted in \cref{fig:MesonBSE}.
%

To solve the vector meson BSEs, we need two ingredients: the quark propagators and the pseudoscalar Bethe-Salpeter amplitudes (BSAs).
The former is obtained from its DSE/EOM depicted in \cref{fig:quarkgap}.
To connect to the commonly used notation for the quark propagators, from now on, we write $S_f$, where $f$ denotes the quark flavour. The inverse propagators are conveniently parameterised by $A_f(p^2)$ and the quark mass function $M_f(p^2)$, 
\begin{align}\label{eq:inverse-quark-prop}
	S_f(p)^{-1} = A_f(p^2) \left[\textrm{i} \slashed{p} + M_f(p^2)\right]\,.
\end{align}
The propagator is diagonal in colour space, and we suppress the respective indices as well as explicit spinor indices. We use Euclidean conventions throughout this work, i.e., ${p^2 = p_0^2 + \vec{p}^2}$ and $\lbrace\gamma_\mu,\gamma_\nu\rbrace = 2 \delta_{\mu\nu}$.

To solve the DSE, the gluon propagator and the quark-gluon vertex are needed.
Using the rainbow truncation, they can be described by an effective interaction.
It enters via the following replacement:
\begin{align}\label{eq:rainbowladder}
	V^{\bar q q A}_\mu(p,q)\, G^{AA}_{\mu\nu}(k)S^{\bar q q A}_\nu \to - Z_2^2\gamma_\mu \, \Pi^T_{\mu\nu}(k) {\mathcal G}(k)\gamma_\nu\,,
\end{align}
where $V^{\bar q q A}_\mu$ and $S^{\bar q q A}_\nu$ are the full and bare quark-gluon vertices, respectively, $G^{AA}_{\mu\nu}$ the gluon propagator, $Z_2$ the quark wave function renormalisation and $\Pi^T_{\mu\nu}$ the transverse projector. The effective interaction ${\mathcal G}(k)$ is a function of the gluon momentum $k$ only, with parameters chosen to reproduce the pion decay constant $f_\pi$.

Throughout this work, we use the Maris-Tandy (MT) model \cite{Maris:1999nt} for the effective interaction ${\mathcal G}(k)$ which has been shown to provide a good description of pseudoscalar and vector mesons as well as baryons and tetraquarks \cite{Cloet:2013jya,Eichmann:2016yit,Eichmann:2020oqt,Hagel:2025ngi}. The details of the model, including the employed parameter values, can be found in \cref{app:MTmodel}. The dependence on this model is the largest systematic uncertainty of the present computation.
The MT model also introduces non-causal structures in the quark propagator \cite{Dorkin:2013rsa,Windisch:2016iud}.
Alternative models such as the one in \cite{Pawlowski:2024kxc} preserve causality and allow for computations arbitrarily deep in the complex plane, but they do not remove the inherent model uncertainty  either.
Since we verified that the non-causal structures are not relevant in the momentum ranges we consider, we continue with the MT model, which is well explored and allows for comparisons with previous computations.
Recent developments to include full quark-gluon vertices in bound state equations can be found in \cite{Ferreira:2025wpu, Ferreira:2026gbe} and calculations of the vertex itself, e.g., in \cite{Williams:2015cvx,Cyrol:2017ewj,Gao:2021wun,Aguilar:2024ciu,Wieland:2026iml}.

With the MT model for the quark-gluon interaction, we can solve the quark gap equation for the light and strange quark propagators $S_f$.  
The equation is discretised on a parabolic momentum grid in the complex plane and solved iteratively. For details, see \cref{app:Numerics-quark}.

The same model is used in the pseudoscalar BSE, for which we employ the ladder approximation.
It contains only the one-gluon exchange diagram, which is directly related to the quark gap equation.
This combination of truncations is called the rainbow-ladder (RL) truncation. It is the simplest way to respect chiral symmetry and guarantees a massless pion in the chiral limit.
We also employ the ladder approximation for the one-gluon exchange in the vector meson BSE.
Starting from the 4PI kernels, we thus modify this diagram accordingly.
Diagrammatically, we represent this by dressing only one vertex instead of two in the BSE, which makes it consistent with the quark gap equation.

From the pseudoscalar BSEs, we extract the second ingredient of the decay diagrams, the pseudoscalar BSAs. Specialising \labelcref{eq:homogeneous_BSE} to the case of mesons, we have
\begin{align}\label{eq:mesonBSE}\nonumber
	\Gamma_i(p,P) &= \int_q  K_{f f'}(p,q,P)\,S_f(-q_-)\,\Gamma_i(q,P)\,S_{f'}(q_+)\,, \\[2ex]
	&= \int_q  \tilde K_{f f'}(p,q,P)\,\Gamma_i(q,P)\,, 
\end{align}
where $q_+=q+\eta P$ and $q_-=-q + (1-\eta) P$ with the momentum partitioning parameter $\eta \in [0,1]$. In the second line, we introduced the shorthand notation $\tilde K$ for the kernel with the quark propagators attached. The BSA $\Gamma$ is written as a function of the relative momentum $p$ of the two quarks and the total momentum $P$ of the state, which is indicated by the subscript $i$. The integral measure is $\int_q = \int\frac{d^4q}{(2\pi)^4}$. The kernel $K_{f f'}$ is given diagrammatically in \cref{fig:MesonBSE} and depends on the flavours of the quark and antiquark legs. With explicit momentum routing, it reads 
\begin{align}\label{eq:Kernelrouting}
	K_{f f'}(p,q,P) = K_{ff'}(p_+,p_-,-q_+,-q_-)\,,
\end{align}
with all momenta defined as incoming. This entails that $P = p_+ + p_- = q_+ + q_-$ and $q=(1-\eta)\,q_+ - \eta\, q_-$ (likewise for $p$), e.g., $q=(q_+ - q_-)/2$ for $\eta=1/2$. In practice, $q_\pm$ and $p_\pm$ carry the same respective momentum fraction $\eta$ of the total momentum. Thus, no total momentum is routed through the t-channel of the scattering kernel.  
The pion and kaon BSEs are solved with the pure RL kernel. This approximation already reproduces the light pseudoscalar phenomenology. Since the lightest two-body channels allowed by quantum numbers, $\pi\sigma$ and $K\sigma$, respectively, are kinematically closed, the corresponding decay diagrams do not generate a width.

The homogeneity of \labelcref{eq:mesonBSE} allows us to treat it as an eigenvalue problem by simply multiplying the left-hand side by $\lambda ( P^2)$, 
\begin{align}\label{eq:mesonBSE-eigenvalue}\nonumber
	\lambda(P^2)\,\Gamma_i(p,P) &= M(p,q,P)\cdot\Gamma_i(q,P) \\[2ex]
	M(p,q,P) &= \int_{\hat{q}_T} \tilde K_{f f'}(p,q,P)\,.
\end{align}
Here, $M$ is the linear operator that acts on the BSA $\Gamma_i$ and that we seek to diagonalise. The integral measure $\int_{\hat q_T}$ is the integral over the angular part of the loop momentum $q$ in the plane transverse to the total momentum $P$.
A solution of \labelcref{eq:mesonBSE}, viz., the BSA of a state of mass $m_i$, is then given by the eigenvector $\Gamma_i$ with eigenvalue $\lambda(-m_i^2) = 1$. Details on the solution strategies for such BSEs can be found in \cite{Eichmann:2016yit,Sanchis-Alepuz:2017jjd,Eichmann:2025wgs,Huber:2025cbd} and references therein; they are briefly discussed in \cref{app:Numerics-BSE}. 

The pseudoscalar BSAs can be expanded in four Lorentz covariant structures,
\begin{align}
	\Gamma_{PS}(p,P)= \gamma^5 \bigl[&E(p,P) - \textrm{i}\slashed{P}\, F(p,P) \nonumber\\[2ex]
	 &{}\hspace*{-.5cm}- \textrm{i}\slashed{p}\,\, G(p,P) + [\slashed{p},\slashed{P}]\, H(p,P)\bigr]\,, \label{eq:ps-tensor-structures}
\end{align}
where the scalar dressing functions $E,F,G,H$ depend on the relative momentum $p$ and the total momentum $P$. For the pion, the third tensor structure receives an additional factor $p\cdot P$, so that $G_\pi$ is even in $z=\hat p\cdot \hat P$. Here and in the following, a hat denotes a normalised four-vector, $\hat k = k/\sqrt{k^2}$.
The kaon is not a charge conjugation eigenstate, and its dressing functions are neither even nor odd in $z$. The extra factor $p\cdot P$ would introduce a $1/z$ singularity in the dressing function $G_K(p,P)$.

The BSA is normalised according to the canonical normalisation condition \cite{Nakanishi:1969ph},
\begin{align}\label{eq:nakanishi}
	&\left(\frac{d\ln\lambda(P^2)}{dP^2}\right)^{-1}_{P^2=-m_i^2} \nonumber\\[2ex]
	&= -N_c\Tr\int_q \overline\Gamma_i(q,P)\,S_f(-q_-)\,\Gamma_i(q,P)\,S_{f'}(q_+)\,,
\end{align}
which is equivalent to the normalisation of \cite{Cutkosky:1964zz}.
Here, $\overline\Gamma(q,P) = C\,\Gamma(-q,-P)^T C^T$ is the charge conjugate amplitude with the charge conjugation matrix $C=\gamma_0\gamma_2$. The trace runs only over Dirac indices. The computed values of the pion and kaon masses and decay constants are listed in \cref{tab:ps-mesons}.
With this normalisation, the pion decay constant, calculated with
\begin{align}\label{eq:fpi}
f_\pi=\textrm{i}\frac{Z_2N_c}{P^2}\Tr\int_q \gamma^5 \slashed{P}\,S(-q_-)\,\Gamma_\pi(q,P)\,S(q_+)\,,
\end{align}
is 130.1\,MeV. We divide this by $\sqrt{2}$ to obtain the conventionally quoted value of $f_\pi = 92.0$\,MeV. In both \labelcref{eq:nakanishi,eq:fpi}, the total momentum $P$ is taken on-shell, the colour trace yields a factor of $N_c=3$, and the flavour trace is normalised to unity.
\begin{table}[tb]
	\begin{ruledtabular}
	\begin{tabular}{lcccc}
		 & $m_\pi$ [MeV] & $f_\pi$ [MeV] & $m_K$ [MeV] & $f_K$ [MeV] \\
		\colrule
		RL (MT)   & 138$^\dagger$   & 92.0$^\dagger$  & 494$^\dagger$ & 108.7 \\
		Exp.      & 138.0 & 92.1  & 493.7 & 110.1 \\
	\end{tabular}
	\end{ruledtabular}
	\caption{Masses and decay constants of the pseudoscalar mesons in the RL truncation with the MT model, compared to experiment~\cite{ParticleDataGroup:2026mpi}. The current quark masses and the coupling strength (of the MT model) are fixed such that $m_\pi$, $f_\pi$ and $m_K$ are reproduced. These values are marked with $^\dagger$. \hspace*{\fill}}
	\label{tab:ps-mesons}
\end{table}

Finally, the evaluation of the decay diagrams at time-like total momenta also requires the pion/kaon BSAs as functions of complex relative momenta. Given a solution of the BSE for Euclidean relative momenta, we exploit the fact that the BSA is analytic in the relative momentum up to the quark propagator poles. As long as the required momentum range does not reach these poles, the form of the RL kernel allows us to evaluate the right-hand side of \labelcref{eq:mesonBSE} for complex relative $p$. The internal momentum $q$ stays on the real Euclidean axis. 

\subsection{Vector mesons and decay diagrams above threshold}\label{sec:vector-mesons}

Having set up the pseudoscalar meson input and the quark propagators for the decay diagrams, we now turn to the vector meson BSE with the full decay kernel. It is given by
\begin{align}\label{eq:Kernelsplit}
	K_{f f'}(p,q,P) = K^{\text{RL}}_{f f'}(p,q,P) + K^{\text{decay}}_{f f'}(p,q,P)\,,
\end{align}
where $K^{\text{RL}}$ is the same rainbow-ladder kernel as for the pseudoscalar mesons and $K^{\text{decay}}$ is the sum of the box and crossed-box diagrams. 
In the following, we focus on the decay kernel, as the rainbow-ladder part of the kernel is treated in the same way as for the pseudoscalar mesons. 

For the decay diagrams, the integrand of the right-hand side of \labelcref{eq:mesonBSE} can be written as 
\begin{align}\label{eq:decay-diagrams}\nonumber
	&\int_q K^{\text{decay}}_{f f'}(p,q,P)\,S_f(-q_-)\,\Gamma_i^\mu(q,P)\,S_{f'}(q_+)\\[2ex]=
	&-\int\limits_{q,l} \!\mathcal{J}_l(p,l,P) \cdot \mathcal{J}^\mu_{r}(q,l,P) D_{+}(l_+)D_{-}(l_-)\,,
\end{align}
where we dropped the explicit flavour indices in the second line. Here, $\mathcal{J}_l$ is the common left part of both the box and the crossed-box diagram, and we refer to the sum of the right triangles as 
 $\mathcal{J}_r^\mu = \mathcal{J}^\mu_{\text{box}} + \mathcal{J}^\mu_{\text{cr}}$. To make the expressions tractable, we use a superindex notation in the following. The full expressions for $\mathcal{J}_{l,r}$ are written out in \cref{app:DecayBSSymmetries}. The left part $\mathcal{J}_{l}$ has open flavour, colour and Dirac indices and is given by 
\begin{align}\label{eq:J_l}
	\mathcal{J}_l = (\Gamma_{-})_{ab}\,S_{bc}\,(\Gamma_{+})_{cd}\,,
\end{align}
Here, $\Gamma_\pm$ and $D_\pm$ are the Bethe-Salpeter amplitudes and propagators of the intermediate states that carry the momenta $l_\pm$, defined analogously to $q_\pm$ and $p_\pm$. The meson propagators are not dressed in this setup; they are simple poles at the masses of the intermediate states,
\begin{align}\label{eq:pion-prop}
	D_\pm(l) = \frac{1}{l^2 + m_\pm^2}\,.
\end{align} 
In this work, we only have $m_\pm \in \lbrace m_\pi, m_K\rbrace$. The pseudoscalar BSAs are evaluated on the mass shell, i.e., at $l_\pm^2 = -m_\pm^2$.

The right part of the decay diagrams is given by the sum of the box and crossed-box diagrams, 
\begin{align}\label{eq:J_right}
	\mathcal{J}^\mu_{\text{box}} &= S_{ab}\,(\Gamma_{i}^\mu)_{bc}\,S_{cd}\,(\Gamma_{+})_{de}\,S_{ef}\,(\Gamma_{-})_{fa}\,,\\[2ex]
	\mathcal{J}^\mu_{\text{cr}} &= S_{ab}\,(\Gamma_{i}^\mu)_{bc}\,S_{cd}\,(\Gamma_{-})_{de}\,S_{ef}\,(\Gamma_{+})_{fa}\,,
\end{align}
where the indices $a,\dots,f$ combine Dirac, colour and flavour indices. Here, the vector meson Bethe-Salpeter amplitude $\Gamma_i^\mu$ enters, with $i \in \lbrace \rho,K^*,\phi\rbrace$. It is transverse to the total momentum $P$ and can be expanded in eight Lorentz covariant structures,	
\begin{align}\label{eq:rho-tensor-structures}
	\Gamma_i^\mu(p,P) = \sum_{n=1}^8 f_i^n(p^2,p\cdot P,P^2)\,\tau_n^\mu(p,P)\,,
\end{align}
with the basis tensors $\tau_n^\mu$ and the scalar dressing functions $f_i^n$. Dirac, colour and flavour indices are suppressed. We use a normalised basis, see, e.g., \cite{Sanchis-Alepuz:2017jjd,Hagel:2025ngi}; the explicit expressions are listed in \cref{app:rho-tensor-structures}. For the $\rho$ and $\phi$ mesons, the dressing functions are even in $p\cdot P$, analogously to the pseudoscalar case. To this end, the two structures $\tau_2^\mu$ and $\tau_7^\mu$, which would otherwise be odd under $p\cdot P\to-p\cdot P$, carry an explicit factor of $z=\hat p\cdot\hat P$. Using this together with the charge conjugation properties of the full amplitudes, we work out the resulting symmetry properties of the decay diagrams in \cref{app:DecayBSSymmetries}. 
\begin{figure*}[ht]
	\includegraphics[width=\textwidth]{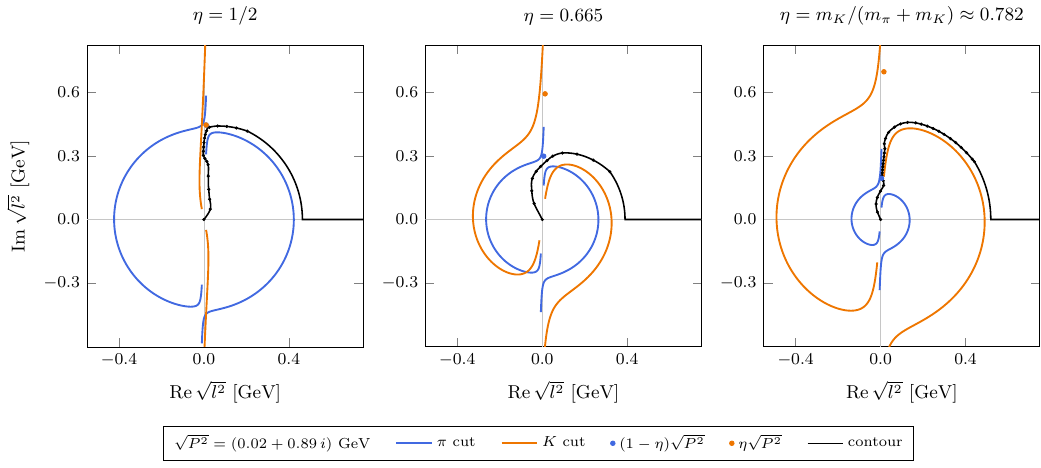}
	\caption{Branch cuts of the pion (blue) and kaon (orange) propagators in the complex $\sqrt{l^2}$ plane and the deformed integration contour (black) for the $K^*$ decay diagram at $\sqrt{P^2} = (0.02 + 0.896\,\textrm{i})$\,GeV, for three choices of the momentum partitioning $\eta$. The dots mark $(1-\eta)\sqrt{P^2}$ and $\eta\sqrt{P^2}$.\hspace*{\fill}}
	\label{fig:unequal-contour}
\end{figure*}

With the Dirac, colour and flavour indices traced out, \labelcref{eq:J_right} is a Lorentz four-vector and can be decomposed into transverse and longitudinal parts with respect to the total momentum $P$. The flavour traces, which include the distinction between $\pi^{0,\pm}$ and $K^{0,\pm}$ for the different processes, are discussed in \cref{app:flavortrace}. Following the discussion in \cite{Williams:2018adr}, the same decomposition can be applied to the left part of the decay diagrams, $\mathcal{J}_l$, after projecting it onto the $n$-th tensor structure by multiplying from the left by the projector $\bar \tau_n^\mu$ and taking the trace. The object to compute is then 
\begin{align}\label{eq:Kernelsplit2}
	M_{nm}^{\text{decay}}  &=  \int_{\hat q_T,l}\!\!L_n^\mu D_{+}(l_+)D_{-}(l_-) R_m^\mu \,, 
\end{align}
where $M^{\text{decay}}$ is defined as the decay part of $M$ in \labelcref{eq:mesonBSE-eigenvalue}. The indices $n$ and $m$ run over the eight tensor structures of the vector meson BSA. The objects $L_n^\mu$ and $R_m^\mu$ are obtained
from the left and right parts of the decay diagrams, 
\begin{align}\label{eq:LR_parts}
	L_n^\mu(p,l,P) = \Tr_{d,c,f}\left\lbrace \bar{\tau}_n^\mu(p,P)\,\mathcal{J}_l(p,l,P)\right\rbrace\,,\\[2ex]
	\mathcal{J}_r^\mu(q,l,P) = R_m^\mu(q,l,P) f_m(q^2,q\cdot P,P^2)\,.
\end{align} 
The second line is an implicit definition of $R_m^\mu$ as the part of $\mathcal{J}_r^\mu$ that carries the $m$-th tensor structure of the vector meson BSA but not the corresponding dressing function. 
Since the integral measure $\int_{\hat q_T}$ is the integral over the angular part of the loop momentum $q$ in the plane transverse to the total momentum $P$, the integral in \labelcref{eq:Kernelsplit2} is six-dimensional in total, of which four are non-trivial, and comes with considerable numerical cost if evaluated as it is. The factorisation in \labelcref{eq:Kernelsplit2}, however, reduces this cost considerably: a particular choice of preferred directions in the angular integrations allows us to factorise the discretised kernel into a product of two matrices, each of which is only a (matrix-valued) one-dimensional integral. The matrix product then represents the remaining two non-trivial integrals. The details of the numerical implementation are discussed in \cref{app:num-factorisation}.

The factorisation of the decay diagrams into left and right parts is not a mere coincidence. The amputated right part of the decay diagram, i.e., $\mathcal{J}_{\text{box/cr}}$, is given by the diagrams of the impulse approximation to the vertex $V\to P_1 P_2$, where $V$ is the vector meson and $P_{1,2}$ are the pseudoscalar mesons. This amplitude has been used to compute the decay width of vector mesons \cite{Jarecke:2002xd,Mader:2011zf}, where the impact of the decay diagrams on the bound state dynamics was neglected. We compare our results with these computations in \cref{sec:literaturecomparison}, where we also make the connection between the BSE solution and the impulse approximation explicit. 

\subsection{Contour deformations}\label{sec:contour-deformations}

To solve the vector meson BSE with the decay diagrams, we need to evaluate the kernel for complex total momenta $P^2$. For the case of equal masses of the decay products, as for the $\rho\to\pi\pi$ and $\phi\to K\bar K$ decay kernels, the threshold for the branch cut in the complex $P^2$ plane is given by $P^2 = -4 m_\pi^2$ and $P^2 = -4 m_K^2$, respectively. A computation at complex momenta with $\operatorname{Re}\, P^2 < -4 m^2$, where $m$ is now either the pion or kaon mass, requires a deformation of the integration contour of the loop momentum $l$ in \labelcref{eq:Kernelsplit2} to be well defined. For detailed discussions, see \cite{Maris:1995ns,Alkofer:2003jj,Windisch:2013dxa,Eichmann:2019dts,Eichmann:2021vnj,Huber:2022nzs,Huber:2023uzd}. This technique has been used, e.g., for the $\rho$ meson \cite{Williams:2018adr}, the quark-photon vertex \cite{Miramontes:2019mco}, and the time-like electromagnetic pion form factor~\cite{Miramontes:2021xgn}. 

The choice of the integration contour is not unique. Our choice maximises the distance to the singularities in the integrand. The details of the contour choice are discussed in \cref{app:contour}. In the case of unequal masses, as for the $K^*\to K\pi$ decay, the cut structure of the integrand increases the numerical difficulty and can require more complicated contour choices. We show three examples for the $K^*$ in the complex $\sqrt{l^2}$ plane in \cref{fig:unequal-contour} where the total momentum $P$ is chosen such that $\operatorname{Im}\, \sqrt{P^2} \approx m_{K^*}$ and the real part is sufficiently close to zero to facilitate the following discussion. The cuts of the pion and kaon propagators are shown in blue and orange, respectively, and mark the movement of the pion and kaon poles as functions of $z_l=\hat l\cdot\hat P \in [-1,1]$. Here, we chose to route $\eta P$ through the kaon line. The black lines show possible deformed integration contours for three different choices of the momentum partitioning $\eta$. The blue and orange dots mark the points $(1-\eta)\sqrt{P^2}$ and $\eta\sqrt{P^2}$. The arguments in \cite{Eichmann:2019dts} generalise to the case of unequal masses and $\sqrt{P^2}$ in the upper-right quadrant, so we can always use a contour that goes in a straight line from the origin to the point $\eta\sqrt{P^2}$ (or $(1-\eta)\sqrt{P^2}$, depending on the distance to the origin), then bends around the cuts in an arc up to the real axis and finally goes in a straight line to infinity. 
Such a contour, however, comes very close to singularities if $\operatorname{Re}\,\sqrt{P^2}$ becomes small. 
Therefore, we choose the contour such that it maximises the smallest distance to the cuts. The algorithm to find such a contour numerically is explained in \cref{app:contour-unequal}. 

As visible in \cref{fig:unequal-contour}, how close a contour has to come to the cuts depends on the choice of the momentum partitioning $\eta$. The first panel shows equal partitioning, $\eta=0.5$. In this case, only the pion cut crosses the real axis, while the kaon cut merely obstructs the contour in the complex plane. In general, the cuts cross the real axis once $\eta\, \operatorname{Im}\, \sqrt{P^2} > m_K$ or $(1-\eta) \operatorname{Im}\, \sqrt{P^2} > m_\pi$. This allows the definition of an ``ideal'' $\eta$ that maximises the range of $\operatorname{Im}\, \sqrt{P^2}$ for which no contour deformation is required. It is given by $\eta = m_K/(m_\pi + m_K) \approx 0.782$ and requires a deformation of the contour only for $\operatorname{Im}\, \sqrt{P^2} > m_\pi + m_K$. It is ideal for computations below threshold but turns out to be a poor choice above it. Instead, we found that for computations in a window around the $K^*$ mass, $\eta \approx 0.665$ is a good choice, as it moves the cuts on top of each other and allows for a contour that has to thread the needle only once. In fact, we can always find such an $\eta$ as the energy fraction of the kaon,
\begin{align}\label{eq:eta-onshell}
	\eta_\text{os}(M) = \frac{M^2 + m_K^2 - m_\pi^2}{2M^2}\,,
\end{align}
where $M$ is the mass or rather the (off-shell) energy of the decaying particle; in our case, $M = \operatorname{Im}\, \sqrt{P^2}$. This construction is not limited to the present case and generalises directly. Note that away from the time-like axis, the matching of the cuts is only approximate.

We checked explicitly that the resulting eigenvalues of the $K^*$ BSE agree within numerical accuracy for different choices of $\eta$ and the corresponding contours. In principle, as the eigenvalues do not depend on the choice of $\eta$, one is not limited to a single momentum partitioning in the construction of the eigenvalue curve, allowing for further numerical optimisations. This closes our discussion of the computational setup for the decay diagrams and the vector meson BSE. 

\section{Results}\label{sec:results}
In the following, we present our results for the $\rho$, $K^*$ and $\phi$ mesons and compare to results from previous calculations.

\subsection{Masses and decay widths for $\rho$, $K^*$ and $\phi$}\label{sec:results_rho_phi_K}

A resonance corresponds to a pole on the second Riemann sheet, located where the eigenvalue of the BSE reaches unity, $\lambda(P^2) = 1$. In the complex $\sqrt{P^2}$ plane, the poles lie at $\sqrt{P^2} = -\Gamma/2 + \textrm{i}M$, from which we read off the mass $M$ and the width $\Gamma$ of the resonance. 

The BSEs, however, are solved only on the first Riemann sheet, for a random set of complex total momenta $P^2$ with $\sqrt{P^2}$ in the upper-right quadrant of the complex plane. For each meson, the sampling regions are chosen such that all points lie safely above the respective decay thresholds. To reach the second sheet, we analytically continue the eigenvalue curve with the Schlessinger point method (SPM) \cite{Schlessinger:1968vsk}. For the first applications of the SPM to the computation of resonance poles, see \cite{Haritan:2017spm,Tripolt:2016cya,Tripolt:2018xeo}. We solve $\lambda(P^2) = 1$ for the SPM continuation of subsets of the sampled points. The sampling, the subsets and the error estimate are described in \cref{app:spm}.
\begin{table}[tb]
	\renewcommand{\arraystretch}{1.3}
	\begin{ruledtabular}
	\begin{tabular}{llccc}
		 & $\pi/K$ BSA & $c$ & $M$ [MeV] & $\Gamma$ [MeV] \\
		\hline\hline
		 & $B/f_\pi$ & 1 & $521.8 \pm 0.9$ & $117.3 \pm 1.7$ \\
		 & $E,F$ & 1 & $622.1 \pm 0.3$ & $60.1 \pm 0.7$ \\
		$\rho$ & $E,F,G,H$ & 1 & $625.1 \pm 0.3$ & $59.9 \pm 0.6$ \\
		 & \boldmath$E,F,G,H$ & \textbf{0.7067} & {\boldmath$774.9 \pm 0.6\rlap{$^\dagger$}$} & {\boldmath$109.1 \pm 1.2$} \\
		 & Exp.~\cite{ParticleDataGroup:2026mpi} &  & $775.26 \pm 0.23$ & $147.4 \pm 0.8$ \\
		 \colrule
		 & $E,F,G,H$ & 1 & $862.4 \pm 0.1$ & $25.1 \pm 0.2$ \\
		$K^*$ & \boldmath$E,F,G,H$ & \textbf{0.9446} & {\boldmath$895.5 \pm 0.1\rlap{$^\dagger$}$} & {\boldmath$31.7 \pm 0.2$} \\
		 & Exp.~\cite{ParticleDataGroup:2026mpi} &  & $895.56 \pm 0.20$ & $47.1 \pm 0.5$ \\
		\colrule
		 & $E,F,G,H$ & 1 & $1003.7 \pm 0.2$ & $1.2 \pm 0.4$ \\
		$\phi$ & \boldmath$E,F,G,H$ & \textbf{0.9726} & {\boldmath$1019.4 \pm 0.1\rlap{$^\dagger$}$} & {\boldmath$2.7 \pm 0.3$} \\
		 & Exp. $\Gamma_{K\bar K}$ ~\cite{ParticleDataGroup:2026mpi} &  & $1019.460 \pm 0.016$ & $3.556 \pm 0.033$ \\
		\end{tabular}
	\end{ruledtabular}
	\caption{Pole positions $\sqrt{P^2} = -\Gamma/2 + \textrm{i}M$ from the SPM for the $\rho$, $K^*$ and $\phi$. For each meson, we list the results for the computed RL scale $c = 1$, for the tuned $c^*$ with $M = M_\text{PDG}$ (bold) and the experimental values ($K^{*0}$ for the $K^*$). Values marked with $^\dagger$ are fixed by the tuning with errors from the SPM extrapolation. For the $\phi$ meson, we compare with the $K\bar K$ partial width.\hspace*{\fill}}
	\label{tab:vm-poles}
\end{table}
\begin{figure*}[ht]
	\includegraphics[width=\textwidth]{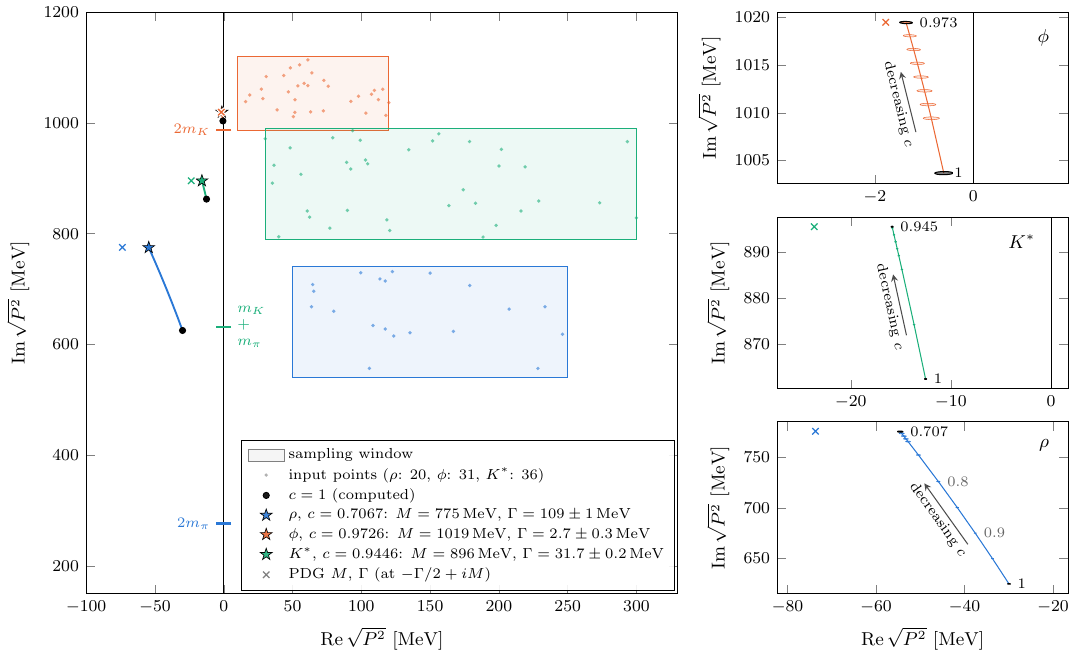}
	\caption{SPM poles of the $\rho$, $K^*$ and $\phi$ in the complex $\sqrt{P^2}$ plane. Left: sampling windows and input points, two-body thresholds on the time-like axis, and the pole trajectories from the computed RL scale $c = 1$ (black dots) to the tuned $c^*$ (stars), together with the PDG values (crosses at $-\Gamma/2 + \textrm{i}M$). Right: zooms on the trajectories with the $1\sigma$ ellipses of the SPM poles; the numbers give $c$. \hspace*{\fill}}
	\label{fig:spm-overview}
\end{figure*}

The results for the pole positions of the $\rho$, $K^*$ and $\phi$ are listed in \cref{tab:vm-poles} and shown in \cref{fig:spm-overview}. We performed the computations for all vector mesons with the full pseudoscalar amplitudes, i.e., with all four tensor structures $E,F,G,H$. For the $\rho$ meson, we also checked the effect of using only the two leading tensor structures $E$ and $F$, as well as the approximation of \cite{Williams:2018adr}. There, the pion BSA was approximated by its leading tensor structure $E$ and the chiral identity \cite{Maris:1997hd}
\begin{align}
	\Gamma_\pi(p,P) \approx \sqrt{2}\frac{B(p^2)}{f_\pi}\,\gamma^5\,,
\end{align}
where $B(p^2) = M_l(p^2)\, A_l(p^2)$ is the scalar dressing function of the inverse light quark propagator \labelcref{eq:inverse-quark-prop}. The factor $\sqrt{2}$ is due to the normalisation of the flavour structure of the pion and the convention $f_\pi=92\,\text{MeV}$. The approximations of the $\pi/K$ BSAs are indicated in the second column of \cref{tab:vm-poles}.

We find that the effect of adding the second tensor structure $F$ to the pion BSA is far from negligible. In contrast to the dressing function $E$, it acts repulsively in the decay kernel and reduces the width of the $\rho$ meson by about $50\%$. This becomes even more evident in the comparison with the impulse approximation in \cref{sec:literaturecomparison}. The remaining two tensor structures $G$ and $H$ act in opposite directions: at the level of the on-shell triangle coupling, adding only $G$ or only $H$ changes $g_{\rho\pi\pi}$ by about $-7\%$ and $+5\%$, respectively, see \cref{sec:literaturecomparison}. Their effects on the width largely cancel, such that the full result differs from the $E,F$ approximation by less than $3\%$. Comparing with the results of \cite{Williams:2018adr}, we find that we can reproduce their results for the $\rho$ meson only if we multiply the decay kernel in \labelcref{eq:Kernelsplit2} by a factor $\frac12$. Given the additional independent cross-checks of the $\rho\to\pi\pi$ coupling in \cref{sec:literaturecomparison}, we are confident that our results are correct.

For the $\rho$ meson, the computed mass is far below the experimental value. This is mainly rooted in the model dependence of the one-gluon exchange. Without the decay diagrams, the MT model reproduces the experimental masses of the vector mesons. Attractive interactions such as the decay diagrams reduce the masses and have to be cancelled in the full theory by either repulsive parts in the one-gluon exchange or additional diagrams.
The masses of the $K^*$ and $\phi$ are much closer to the experimental values, as their widths and hence the effects of the decay diagrams are much smaller. In particular, the $\phi$ is so close to the $K\bar K$ threshold that the decays are strongly phase-space suppressed. 

For the $\rho$ and $K^*$ mesons, the decay diagrams we compute are the only diagrams that contribute significantly to the decay width. For the $\phi$ meson, there is also the open $\phi \to \pi\pi\pi$ channel, which carries approximately $15\%$ of the total width and is not contained in our setup. It appears only at two-loop order in the kernel and is beyond the scope of this work. It could be included by considering sequential decays via the channel $\phi \to \rho \pi$ and including a dressed propagator for the $\rho$. Effectively, this accounts for the $\pi\pi$ threshold currently stored in (and dropped with) the rest term of the on-shell decomposition \labelcref{eq:4q-onshell-decomposition}. Similar arguments apply to higher numbers of pions in the final state, which we do not account for: the $\phi$ can potentially decay to up to $7$ pions. For the $\rho$ and $K^*$ mesons, only the $\rho \to 4\pi$ and $K^*\to K\pi\pi$ channels are open and allowed in addition to the ones considered. However, these processes are heavily suppressed and contribute to the respective widths only at the sub-permille level. Other diagrams (of strong decays) cannot contribute to the width of the considered vector mesons, as potential decays into quarks or gluons are forbidden by confinement and decays into other hadrons are kinematically closed. These diagrams merely shift the mass parameter $M$ of the resonance.

In terms of analyticity, the branch cuts of the open channels manifest as imaginary parts of the eigenvalue curve on the time-like axis,
\begin{align}\label{eq:disc}
	\text{disc}\, \lambda(P^2) =  2\textrm{i}\,\operatorname{Im}\, \lambda(P^2 + \textrm{i}\epsilon)\,.
\end{align}
Excluding the strongly suppressed multi-pion processes, the statement above translates to:
\textit{For the considered vector mesons, the sum of all missing diagrams only shifts the real part of the eigenvalue curve on the time-like axis, while the discontinuities are determined by the decay diagrams we compute. }
The only model dependence of these cuts lies in that of the RL ingredients, namely the pseudoscalar meson BSAs and the quark propagators. This is not the case for the widths, as the attractive interactions of the decay diagrams reduce the mass of the resonance and hence reduce the allowed phase space for the decay. This attractive effect is counteracted by further repulsive interactions in the full theory.
To reduce this model dependence of the width, we introduce a tuning parameter in the kernel \labelcref{eq:Kernelsplit}, which then reads 
\begin{align}\label{eq:Kernelsplit-tuned}
	K_{f f'}^{c}(p,q,P) = c \,K^{\text{RL}}_{f f'}(p,q,P) + K^{\text{decay}}_{f f'}(p,q,P)\,.
\end{align}
The tuning parameter $c$ is a real number that adjusts the strength of the one-gluon exchange to reproduce the experimental masses. This amounts to a shift of $\operatorname{Re}\, \lambda(P^2)$ that accounts for the missing diagrams and repulsive interactions while keeping the cut structure introduced by the decay diagrams fixed. Note that this tuning affects the magnitude of the discontinuity \labelcref{eq:disc} only at second order via the iteration of the BSE.

The tuned values of $c$ for the three vector mesons, together with the corresponding masses and the resulting widths, are listed in bold in \cref{tab:vm-poles} and compared to the experimental values listed below them. For the $\phi$ meson, we compare with the $K\bar K$ partial width $\Gamma_{K\bar K} = \Gamma(K^+K^-) + \Gamma(K_L K_S)$~\cite{ParticleDataGroup:2026mpi}, since the $3\pi$ decays of the $\phi$ are not included in our decay diagrams. For the $\rho$ and the $K^*$ mesons, the two-meson channels dominate the full strong interaction part of the width. The tuned values of $c$ are all below unity, which is consistent with the expectation that the decay diagrams are attractive in total. For the $\rho$ meson, the necessary tuning is strongest, which is a consequence of it being composed of light quarks and decaying into the lightest pseudoscalar mesons. Moreover, the $\rho$ meson is considerably above the $\pi\pi$ threshold, leading to a large available phase space for the decay and a correspondingly large effect of the decay diagrams. For the $K^*$ meson, the coupling to the decay channel is suppressed by the larger masses of both the strange quark and the kaon. The smallest necessary tuning is found for the $\phi$ meson, mostly due to its proximity to the $K\bar K$ threshold. 

The tuning is also visualised in \cref{fig:spm-overview}, where we show the trajectories of the poles in the complex $\sqrt{P^2}$ plane from the computed RL scale $c = 1$ to the tuned $c^*$, where $M(c^*) = M_\text{PDG}$. The thresholds on the time-like axis are indicated by small horizontal lines. The right panels zoom in on the $c$ trajectories. The ellipses are the $1\sigma$ errors of the extrapolated resonance poles. The widths of the tuned $\rho$ and $K^*$ mesons are about $26\%$ and $33\%$ smaller than the experimental values, and that of the $\phi$ meson is about $24\%$ smaller than its $K\bar K$ partial width. Deviations of this size are readily explained by the remaining model dependence of the RL ingredients, which we cannot account for in the present setup. For example, it is known that if a full quark-gluon vertex is used instead of the RL approximation, both the mass function of the quark propagator and the pseudoscalar meson BSAs change. The quark mass function decreases in the infrared compared to the RL case, which leads to an enhancement of the decay diagrams. We therefore expect that going beyond RL constitutes an important step towards quantitative agreement with experiment. We also stress that the resolution of a resonance as narrow as the $\phi$ meson is a non-trivial achievement of our numerical setup. 

\subsection{Connection to impulse approximation}\label{sec:literaturecomparison}

In the following, we compare our results with the impulse approximation and draw the connection to the BSE by treating the decay diagrams as a perturbation to the RL kernel. 

Following the discussions in \cite{Jarecke:2002xd,Mader:2011zf}, the transition matrix element of the decay $V\to P_1 P_2$ is given by the triangle diagrams encoded in the right part of our decay diagrams. It defines the coupling $g$, 
\begin{align}\label{eq:triangle-coupling}
	\Lambda^\mu(l,P)\bigg|_\text{os} = \int_q \mathcal{J}_r^\mu(q,l,P) \bigg|_\text{os}= 2g\,l_T^\mu,\quad l_T^\mu= \Pi^T_{\mu\nu}(P)\,l_\nu\,,
\end{align}
where the subscript ``os'' indicates that the external momenta are on shell, i.e., $P^2 = -M_V^2$ and $l_{1,2}^2= -M_{1,2}^2$  with $l_1 = l_+$ and $l_2 = l_-$. This coupling can be related to the decay width of a vector meson via the narrow-width approximation, 
\begin{align}\label{eq:narrow-width}
	\Gamma_{V\to P_1 P_2} = \frac{g^2}{48\pi M_V^5}\,\lambda^{3/2}(M_V^2,M_1^2,M_2^2)\,,	
\end{align}
where $\lambda(x,y,z) = x^2 + y^2 + z^2 - 2xy - 2xz - 2yz$ is the K\"all\'en function.  
\begin{table}[tbp]
	\renewcommand{\arraystretch}{1.3}
	\begin{ruledtabular}
	\begin{tabular}{llccc}
		 & $\pi/K$ BSA & $g$ triangle & $g$ first order & $g$ exp.~\cite{ParticleDataGroup:2026mpi} \\
		\hline\hline
		 & $B/f_\pi$ & $9.081$ & $9.081 \pm 0.023$ &  \\
		$\rho$ & $E,F$ & $5.278$ & $5.279 \pm 0.005$ & $5.94 \pm 0.02$ \\
		 & $E,F,G,H$ & $5.170$ & $5.170 \pm 0.005$ &  \\
		\colrule
		$K^*$ & $E,F,G,H$ & $4.830$ & $4.826 \pm 0.015$ & $5.39 \pm 0.03$ \\
		\colrule
		$\phi$ & $E,F,G,H$ & $4.165$ & $4.159 \pm 0.009$ & $4.53 \pm 0.02$ \\
	\end{tabular}
	\end{ruledtabular}
	\caption{Couplings $g_{\rho\pi\pi}$, $g_{K^*K\pi}$, which combines the charged and neutral channels ($g_{K^*K\pi}^2 = \tfrac32\,g_{K^{*0}K^+\pi^-}^2$), and $g_{\phi K^+K^-}$ from the on-shell quark triangle \labelcref{eq:triangle-coupling} and from the first-order perturbation \labelcref{eq:width-first-order} of the BSE eigenvalue by the decay diagrams. The experimental couplings follow from the PDG widths in \cref{tab:vm-poles} via \labelcref{eq:narrow-width} (for the $\phi$ from $\Gamma_{K\bar K}$), with the physical masses of the charged and neutral decay channels.\hspace*{\fill}}
	\label{tab:vm-couplings}
\end{table}

The evaluation of the triangle diagrams is carried out with the same approximations as for the full decay diagrams. The additional ingredient is the vector meson BSA, for which we use the RL solution evaluated at the on-shell momentum $P^2 = -M_V^2$.
The resulting couplings are listed in the third column of \cref{tab:vm-couplings}. For the $\rho$ meson, the triangle with the full pion amplitudes, $g_{\rho\pi\pi} = 5.170$, agrees with the independent RL triangle result $g_{\rho\pi\pi} = 5.20$ of \cite{Mader:2011zf}. The others are compatible with the results of \cite{Jarecke:2002xd}, where the amplitudes and propagators were continued to the complex plane by Taylor expansion. For the $K^*$ meson, we computed an effective coupling $g_{K^*K\pi}$ that subsumes both decay channels $K^{*0}\to K^+ \pi^-$ and $K^{*0}\to K^0 \pi^0$. In the isospin-symmetric approximation, these channels differ only in their flavour coefficient, see \cref{app:flavortrace}. For the $\phi$ meson, the channels $K^+K^-$ and $K^0\bar K^0$ have the same flavour factor, so their couplings are equal, and we quote $g_{\phi K^+K^-}$ to allow a direct comparison with \cite{Jarecke:2002xd}. 
Compared to the experimental couplings extracted from the PDG widths, the triangle couplings with the full pseudoscalar amplitudes are about $13\%$, $10\%$ and $8\%$ smaller for the $\rho$, $K^*$ and $\phi$, respectively, consistent with the smaller widths of the tuned resonances in \cref{tab:vm-poles}. 
To disentangle the role of the subleading pion tensor structures, we also computed the $\rho$ triangle with only one of $G$ and $H$ added to the $E,F$ amplitude. We find $g_{\rho\pi\pi} = 4.906$ for $E,F,G$ and $g_{\rho\pi\pi} = 5.569$ for $E,F,H$, i.e., changes of $-7.0\%$ and $+5.5\%$ relative to the $E,F$ result. The two structures thus act in opposite directions and largely cancel in the full result, $g_{\rho\pi\pi} = 5.170$. 

The fourth column of \cref{tab:vm-couplings} acts as a cross-check of our decay kernel computation. Since the triangle is evaluated with RL ingredients only, without feeding back the decay diagrams, it corresponds to the first-order effect of the decay diagrams on the RL eigenvalue curve. To see this, we treat the decay part of the kernel in \labelcref{eq:Kernelsplit} as a perturbation to the RL kernel and expand the eigenvalue curve $\lambda(P^2)$ to first order in the decay kernel: 
\begin{align}\nonumber
	&\lambda(P^2) = \lambda_\text{RL}(P^2) + \delta\lambda(P^2) + \mathcal{O}(\delta K^2)\,,\\[2ex]
	&\delta\lambda(P^2) = \frac{\langle \overline{\Gamma}^0_V | K^\text{decay} | \Gamma^0_V\rangle}{\langle \overline{\Gamma}^0_V | \Gamma^0_V\rangle}\,,\label{eq:lambda-first-order}
\end{align}
where the superscript $^0$ indicates the RL solution for the vector meson BSA, and we adopt a suggestive bra-ket notation. We compute this eigenvalue shift from the same kernel matrices that were used in the full BSE computation in \cref{sec:results_rho_phi_K}. The numerator of the second line is nothing but two triangle diagrams, connected by the two pseudoscalar propagators, and the denominator is a normalisation factor. In this sense, $\delta\lambda$ is of order $\mathcal{O}(g^2)$ in the coupling. To compare with the narrow-width approximation, we strictly keep this order in $g$. The corresponding shift of the resonance pole and the resulting first-order width are then given by 
\begin{align}\label{eq:width-first-order}
\delta P^2 = -\frac{\delta \lambda}{\lambda'_\text{RL}}\,,\qquad
\Gamma_{g^2} = -\,\frac{\operatorname{Im}\, \delta P^2}{M_\text{RL}}\,,
\end{align}
where we demanded that $\lambda_\text{RL}(P^2) + \delta\lambda = 1$  and expanded $\lambda_\text{RL}(P^2)$ around the RL pole $P^2_\text{RL} = -M_\text{RL}^2$.
To obtain the width to order $\mathcal{O}(g^2)$, we linearised the parametrisation of the resonance pole $P^2 = -M^2 - \textrm{i}M\Gamma + \Gamma^2/4$ in $\Gamma$, which is itself already of order $\mathcal{O}(g^2)$, in accordance with the narrow-width approximation. Moreover, it is the RL value of the mass that enters the width in \labelcref{eq:width-first-order}, as corrections would be of order $\mathcal{O}(g^4)$ to $\Gamma_{g^2}$. The resulting width is hence of order $\mathcal{O}(g^2)$, and we can use it to compute the coupling $g$ via \labelcref{eq:narrow-width}. Evidently, the resulting couplings are in excellent agreement with the triangle couplings, see \cref{tab:vm-couplings}. The errors originate in the analytic continuation as for the pole positions, see \cref{app:spm}, and give a measure of its stability. This is a non-trivial cross-check of our implementation and demonstrates its internal consistency. Together with the benchmark against literature results, this constitutes a strong validation of our framework. 


\section{Conclusion}\label{sec:conclusion}

In this work, we have presented a framework to implement hadronic decays in (two-body) bound state equations. This includes both the diagrammatic derivation and the numerical implementation. We have applied this framework to the decays of the light vector mesons $\rho$, $K^*$ and $\phi$ into two pseudoscalar mesons and provided results for the respective widths. 
In doing so, we correct and improve the work of \cite{Williams:2018adr} for the $\rho$ meson by using the full pseudoscalar BSAs and extend it to the $K^*$ and $\phi$ mesons.
We showed that at least the two dominant tensor structures of the pseudoscalar Yukawa interactions are necessary, and we included all four of them in the full computation. In all cases, we account for the model dependence of the masses by tuning the strength of the binding interaction to reproduce the experimental masses of the vector mesons, which is particularly relevant for the $\rho$ meson. The remaining deviations of the widths from experiment, around $25$--$33\%$, can be explained by the model dependence of the RL ingredients of the decay diagrams. With the recent advances in the computation of meson BSAs beyond RL, e.g., in \cite{Ferreira:2025wpu}, we expect that this systematic error can be reduced in the near future. 

The derivation of the decay diagrams on the basis of the 4PI effective action of QCD is systematic and has been implemented algorithmically. To this end, we constructed the four-loop 4PI effective action of QCD in \cref{sec:4PI_QCD}, which, to our knowledge, is presented here for the first time. The emerging kernel diagrams contain bubble diagrams with full four-point functions carrying the pole structure of the physical bound states and resonances. The explicit diagrams containing the hadronic decay products in the BSE kernel are algorithmically derived by inserting the on-shell resonance poles in place of the four-point functions, see \cref{sec:decaydiags}. This is not limited to the case of mesons but can be applied to other hadrons as well, e.g., glueballs or baryons. The step to baryons as decay products is diagrammatically clear, but their calculation is complicated by the required six-point functions. 

On the numerical level, the setup based on contour deformations in the complex plane is not specific to the vector mesons considered here: we have provided a general algorithm to find suitable integration contours for the loop integrals of arbitrary two-body decay diagrams.

We find that the tensor structures of the pseudoscalar amplitudes play a central role in the decays. The two dominant ones, $E$ and $F$, are necessary to obtain qualitatively correct results: in the case of the $\rho$ meson, the inclusion of $F$ reduces the width by about $50\%$. The remaining two, $G$ and $H$, each change the coupling by $5$--$7\%$, but with opposite signs; since these contributions largely cancel, they have to be included together for quantitative predictions. We provided independent cross-checks of our results by comparing the decay diagrams with the impulse approximation results. We showed that, on the basis of an existing RL solution, the triangle couplings $g_{VPP}$ are equivalent to the leading-order perturbation of the RL eigenvalue curve by the decay diagrams. This equivalence is not limited to the RL approximation.

In summary, our framework is not limited to the light vector mesons but applies to a wide range of hadronic two-body decays. Its predictive power is currently limited only by the model dependence of the ingredients of the decay diagram and by their analytic structure. Both issues are being addressed in the field, and we hope to report on corresponding results in the near future.

\begin{acknowledgments}
We thank Gernot Eichmann, Christian S. Fischer, Simon Klein, Angel S. Miramontes, H\`elios Sanchis-Alepuz, Richard Williams, and Jonathan Yigzaw for fruitful discussions and work on related projects. This work was supported by the Deutsche Forschungsgemeinschaft (DFG, German Research Foundation) under grant number HU 2176/3-1. It contributes to the aims of the U.S. Department of Energy ExoHad Topical Collaboration, contract DE-SC0023598. We acknowledge computational resources provided by the HRZ of the Justus Liebig University Giessen.

\end{acknowledgments}

\appendix


\section{Vector meson basis}\label{app:rho-tensor-structures}
The basis tensors $\tau_n^\mu(p,P)$ in \labelcref{eq:rho-tensor-structures} depend only on the directions of $p$ and $P$. With $\hat P = P/\sqrt{P^2}$, the normalised transverse relative momentum
\begin{align}
	n^\mu = \frac{p_T^\mu}{\sqrt{p_T^2}}\,,\qquad p_T = p - \frac{p\cdot P}{P^2}\,P\,,
\end{align}
the angle $z = \hat p\cdot\hat P$, and the doubly transverse Dirac matrix
\begin{align}
	\gamma_{TT}^\mu = \gamma^\mu - \hat{\slashed{P}}\,\hat P^\mu - \slashed{n}\,n^\mu\,,
\end{align}
they read
\begin{align}
	\tau_1^\mu &= \tfrac{\textrm{i}}{2}\,n^\mu\,, &
	\tau_5^\mu &= \tfrac{\sqrt2}{4}\,\gamma_{TT}^\mu\,,\nonumber\\
	\tau_2^\mu &= \tfrac{1}{2}\,z\,\hat{\slashed{P}}\,n^\mu\,, &
	\tau_6^\mu &= \tfrac{\sqrt2}{4}\,\hat{\slashed{P}}\,\gamma_{TT}^\mu\,,\nonumber\\
	\tau_3^\mu &= \tfrac{1}{2}\,\slashed{n}\,n^\mu\,, &
	\tau_7^\mu &= \tfrac{\sqrt2}{4}\,z\,\slashed{n}\,\gamma_{TT}^\mu\,,\nonumber\\
	\tau_4^\mu &= \tfrac{1}{2}\,\hat{\slashed{P}}\slashed{n}\,n^\mu\,, &
	\tau_8^\mu &= \tfrac{\sqrt2}{4}\,\textrm{i}\,\hat{\slashed{P}}\slashed{n}\,\gamma_{TT}^\mu\,,
\end{align}
each multiplied by the identity in colour space. All structures are transverse to $P$. The factors of $z$ in $\tau_2^\mu$ and $\tau_7^\mu$ ensure that the dressing functions of the $\rho$ and $\phi$ are even in $p\cdot P$.

\begin{widetext}
%
\section{Structural symmetries of the decay BSE}\label{app:DecayBSSymmetries}

In this appendix, we give the full expressions for the left and right parts of the decay diagrams in \labelcref{eq:decay-diagrams} and discuss their symmetry properties. The left part has open Dirac, colour and flavour indices and reads

\begin{align}\label{eq:J_l_full}
	\mathcal{J}_l(p,l,P) = \Gamma_{-}\Bigl(\frac{(p-l)-p_-}{2},-l_-\Bigr)\, S(p-l)\,\Gamma_{+}\Bigl(\frac{p_+ + (p-l)}{2},-l_+\Bigr)\,,
\end{align}
while the right parts are Lorentz four-vectors,
\begin{subequations}
\label{eq:J_box}
\begin{align}
	&\mathcal{J}^\mu_{\text{box}}(q,l,P)=\Tr_{d,c,f} \left\lbrace
	S(q_+)
	\Gamma_{+}\Bigl(\frac{q-l + q_+}{2},l_+\Bigr)
	S(q-l)
	\Gamma_{-}\Bigl(\frac{q-l - q_-}{2},l_-\Bigr) 
	S(-q_-) 
	\Gamma_{i}^\mu(q,P) 
	\right \rbrace\,,\\[2ex]
	&\mathcal{J}^\mu_{\text{cr}}(q,l,P)=\Tr_{d,c,f} \left\lbrace 
	S(q_+)
	\Gamma_{-}\Bigl(\frac{q+l + q_+}{2},l_-\Bigr)
	S(q+l) 
	\Gamma_{+}\Bigl(\frac{q+l - q_-}{2},l_+\Bigr)
	S(-q_-) 
	\Gamma_{i}^\mu(q,P)
	\right \rbrace\,.
\end{align}
\end{subequations}
For charge conjugation eigenstates, or rather for states whose BSE is even in $q\cdot P$, the box and crossed-box diagrams are related by a flip of the loop momentum $q\to -q$. This is the case for the $\rho$ and $\phi$ mesons. Then, only the symmetric parts of the kernel contribute to the BSEs, which allows us to halve the numerical effort. To see this, we flip the loop momentum $q\to -q$ in the crossed-box diagram, which for $\eta = 1/2$ exchanges $q_+ \leftrightarrow q_-$, and use the charge conjugation properties of the propagators and amplitudes,
\begin{align}\label{eq:chargeconjugation}
	&S(-q) = C\,S(q)^T\,C^{T}\,,\nonumber\\
	&\Gamma_\rho^\mu(-q,P) = \overline{\Gamma}_\rho^\mu(-q,-P) = -C\,\Gamma_\rho^\mu(q,P)^T\,C^{T}\,,\nonumber\\
	&\Gamma_\pi(-q,P) = \overline{\Gamma}_\pi(-q,-P) = C\,\Gamma_\pi(q,P)^T\,C^{T}\,,
\end{align}
where the bar denotes the charge conjugate amplitude and $C$ is the charge conjugation matrix. Note that these relations hold for the full amplitudes; whether the individual dressing functions are even in $q\cdot P$ depends on the choice of the tensor basis. With these relations, we find
\begin{align}
	&\mathcal{J}^\mu_{\text{cr}}(-q,l,P)=\Tr_{d,c} \left\lbrace  
	S(q_-)
	\Gamma_{\pi}\Bigl(\frac{l -q+ q_-}{2},l_-\Bigr)
	S(l-q) 
	\Gamma_{\pi}\Bigl(\frac{l -q - q_+}{2},l_+\Bigr)
	S(-q_+) 
	\Gamma_{\rho}^\mu(-q,P)\right \rbrace\nonumber\\[2ex]
	&=\Tr_{d,c} \left\lbrace 
	C\, S(-q_-)^T \, C^{T}\, 
	\Gamma_{\pi}\Bigl(\frac{l-q + q_-}{2},l_-\Bigr)
	C \,  S(q-l)^T \, C^{T} \,
	\Gamma_{\pi}\Bigl(\frac{-q+l - q_+}{2},l_+\Bigr) \, 
	C \, S(q_+)^T\,C^{T} \,
	\Gamma_{\rho}^\mu(-q,P)\, 
	\right \rbrace\nonumber \\[2ex]
	&=-\Tr_{d,c} \left\lbrace 
	S(-q_-)^T 
	\Gamma_{\pi}\Bigl(\frac{q-l - q_-}{2},l_-\Bigr)^T
	S(q-l)^T 
	\Gamma_{\pi}\Bigl(\frac{q-l + q_+}{2},l_+\Bigr)^T  
	S(q_+)^T 
	\Gamma_{\rho}^\mu(q,P)^T 
	\right \rbrace\nonumber \\[2ex]
	&= - \mathcal{J}^\mu_{\text{box}}(q,l,P)\,,
\end{align}
where we used the cyclicity of the trace twice, $(AB)^T= B^T A^T$ and $\Tr A^T = \Tr A$. 
Hence, in a single-flavour world and for a neutral $\rho$ meson and neutral intermediate pions, the sum of the box and crossed-box diagrams is antisymmetric in $q$, and its total contribution to the decay of this single-flavour $\rho$ vanishes. This is the diagrammatic manifestation of angular momentum conservation and Bose-Einstein statistics: the two-pion final state is in a p-wave and hence has to be antisymmetric under the exchange of the two pions. This is only possible if they are distinguishable, viz., charged. The flavour structure of the charged two-pion state is discussed in \cref{app:flavortrace}.
\end{widetext}
%

\section{Maris-Tandy model}\label{app:MTmodel}
In the RL truncation \labelcref{eq:rainbowladder}, the effective interaction is written as
\begin{align}\label{eq:MT-G}
	{\mathcal G}(k) = \frac{4\pi\,\alpha(k^2)}{k^2}\,.
\end{align}
We use the Maris-Tandy interaction \cite{Maris:1999nt} in the form of \cite{Eichmann:2009qa},
\begin{align}\label{eq:MT-alpha}
	\alpha(k^2) = &\,\pi\,\eta^7 x^2\, e^{-\eta^2 x} \nonumber\\[2ex]
	&+ \frac{2\pi\gamma_m\bigl(1-e^{-k^2/\Lambda_t^2}\bigr)}{\ln\bigl[e^2-1+\bigl(1+k^2/\Lambda_\text{QCD}^2\bigr)^2\bigr]}\,,
\end{align}
with $x=k^2/\Lambda^2$. The first term carries the infrared strength, the second term reproduces the one-loop running of the coupling in the ultraviolet. We use $\Lambda = 0.72$\,GeV and $\eta = 1.8$, together with $\Lambda_\text{QCD} = 0.234$\,GeV, $\gamma_m = 12/25$ and $\Lambda_t = 1$\,GeV. The same parameters are used for all flavours. The quark propagators are renormalised at $\mu = 19$\,GeV with $A_f(\mu^2) = 1$ and $M_f(\mu^2) = m_f(\mu)$, where $m_{u}(\mu) = m_{d}(\mu) = 3.7$\,MeV and $m_s(\mu) = 85$\,MeV.

\section{Flavour Traces}\label{app:flavortrace}
In this appendix, we give the flavour factors of the decay diagrams in \cref{fig:MesonBSE}. Since no quark propagator connects the left and right parts of the decay diagrams, the flavour trace factorises into one trace per quark line. With flavour matrices $F$ normalised to $\tr[F_A F_B^\dagger] = \delta_{AB}$, the two orderings of the pseudoscalar mesons $P_1 P_2$ of a given channel along the quark line of the right part are
\begin{align}
	A = \tr\left[F_V F_{P_1} F_{P_2}\right]\,,\qquad B = \tr\left[F_V F_{P_2} F_{P_1}\right]\,.
\end{align}
We consider both orderings for each quark line. Decays that are not realised in this way vanish in the flavour trace, see \cref{tab:flavour-factors}.
The left part, projected onto the same vector meson, yields the complex conjugates $A^*$ and $B^*$. The box diagram connects each meson vertex on the right with the corresponding vertex on the left and hence contracts each ordering with its own conjugate, while the crossed box exchanges the two meson lines and contracts each ordering with the conjugate of the other. This yields the flavour factors
\begin{align}
	C_\text{box} = |A|^2 + |B|^2\,,\qquad C_\text{cr} = 2\,\operatorname{Re}\,(A B^*)\,.
\end{align}
The resulting factors for the states considered here are listed in \cref{tab:flavour-factors}.
\begin{table}[tb]
	\renewcommand{\arraystretch}{1.3}
	\begin{ruledtabular}
	\begin{tabular}{llcccc}
		$V$ & $P_1P_2$ & $A$ & $B$ & $C_\text{box}$ & $C_\text{cr}$ \\
		\hline\hline
		$\rho^0 = (u\bar u - d\bar d)/\sqrt{2}$ & $\pi^+\pi^-$ & $1/\sqrt{2}$ & $-1/\sqrt{2}$ & $1$ & $-1$ \\
		 & $\pi^0\pi^0$ & $0$ & $0$ & $0$ & $0$ \\
		\colrule
		$K^{*0} = d\bar s$ & $K^+\pi^-$ & $0$ & $1$ & $1$ & $0$ \\
		 & $K^0\pi^0$ & $0$ & $-1/\sqrt{2}$ & $1/2$ & $0$ \\
		 \colrule
		$\phi = s\bar s$ & $K^+K^-$ & $0$ & $1$ & $1$ & $0$ \\
		 & $K^0\bar K^0$ & $0$ & $1$ & $1$ & $0$ \\
	\end{tabular}
	\end{ruledtabular}
	\caption{Flavour orderings $A$, $B$ and flavour factors of the box and crossed-box diagrams for the decay channels considered in this work.\hspace*{\fill}}
	\label{tab:flavour-factors}
\end{table}

\section{Contour deformation}\label{app:contour}
As discussed in \cref{sec:contour-deformations}, above the two-body threshold the decay diagrams require a deformation of the integration contour. Only the radial variable $\sqrt{l^2}$ of the meson loop momentum $l$ is deformed, while the quark momentum $q$ and all angular variables stay real. In this appendix, we discuss the cuts and the choice of the momentum partitioning. For the contour, we distinguish decays into states with equal masses ($\rho\to\pi\pi$, $\phi\to K\bar K$) and unequal masses ($K^*\to K\pi$).

The two decay products carry the momenta $l_+ = l+\eta P$ and $l_- = -l+(1-\eta)P$. For the $K^*$ meson, $l_+$ is the kaon and $l_-$ the pion momentum. With $\eta_+=\eta$ and $\eta_- = 1-\eta$, the poles of the propagator of meson $i$ with mass $m_i$ produce a pair of cuts in the complex $\sqrt{l^2}$ plane,
\begin{align}\label{eq:cuts}
	w_\pm^{(i)}(z_l) = \eta_i\sqrt{P^2}\,z_l \pm \textrm{i}\sqrt{\eta_i^2\, P^2\,(1-z_l^2) + m_i^2}\,,
\end{align}
where $z_l=\hat l\cdot\hat P\in[-1,1]$ is the angular variable.
These cuts directly create the thresholds, see \cite{Windisch:2013dxa,Huber:2023uzd} for details on this mechanism. Note that the subscript $\pm$ here refers to the two cuts and not the two mesons, i.e., there are four cuts in total, see \cref{fig:unequal-contour}. The cuts of meson $i$ cross the real axis when $\eta_i\operatorname{Im}\sqrt{P^2}>m_i$. They leave a gap of half-width $\sqrt{\eta_i^2 M^2 - m_i^2}$ at small $\operatorname{Re}\sqrt{P^2}$, where $M=\operatorname{Im}\sqrt{P^2}$. The two-body threshold lies at the sum of the masses of the two mesons, $M = M_s = m_1 + m_2$.

The integral does not depend on $\eta$, but the geometry of the cuts does. For equal masses, the ideal and the on-shell value of $\eta$ from \cref{sec:contour-deformations} both give $\eta=1/2$, and the two pairs of cuts coincide. For the $K^*$ meson, we use $\eta=\eta_\text{os}(826\,\text{MeV})=0.665$ from \labelcref{eq:eta-onshell}, which keeps the strange quark momentum $q_+$ at $\operatorname{Im}\sqrt{q_+^2}\lesssim 0.6$\,GeV.
This is inside the region where its propagator is analytic, see \cref{app:Numerics-quark}. This single value is used for all $P^2$ in the sampling window and was found empirically to yield suitable contours for all of them.

\subsection{Equal masses}\label{app:contour-equal}

For equal masses and $\eta=1/2$, both mesons produce the same pair of cuts $w_\pm\equiv w_\pm^{(i)}$ of \labelcref{eq:cuts}. For $\sqrt{P^2}$ in the upper-right quadrant, they leave a gap near $z_l\sim 0.6$--$1$, and the contour passes it at the midpoint of the closest approach of the two cuts,
\begin{align}
    w_M &= \frac{w_+(z_1^*)+w_-(z_2^*)}{2}\,,\\[2ex]
    (z_1^*,z_2^*) &= \underset{z_1,z_2}{\arg\min}\,\bigl|w_+(z_1)-w_-(z_2)\bigr|\,,
\end{align}
with $z_1,z_2\in[0,1]$. $z_1^*$ and $z_2^*$ are determined numerically. To keep away from the cuts past $w_M$, the contour is chosen parallel to the cut tangent $T=\partial_{z_l} w_+|_{z_1^*}$. With $R_M=|w_M|$, $\theta_M=\arg w_M$ and $\alpha=\arg T - \theta_M$ (shifted by $\pi$ if $\cos\alpha<0$), this fixes the polar velocities at $w_M$,
\begin{align}
    V_r = R_M\cos\alpha\,,\qquad V_\theta = \sin\alpha\,.
\end{align}
The contour $\gamma(s)=r(s)\,e^{\textrm{i}\theta(s)}$, $s\in[0,2]$, is defined piecewise by segment~1 ($0\le s\le 1$) and segment~2 ($1<s\le 2$):
\begin{align}
    r &= (V_r - R_M)\,s^2 + (2R_M - V_r)\,s\,, \notag\\
    \theta &= \theta_M + V_\theta\,(s-1)\,,\\[2ex]
    r &= R_M + V_r\,\sigma + (L_\text{real}-R_M-V_r)\,\sigma^2\,, \notag\\
    \theta &= \theta_M + V_\theta\,\sigma - (\theta_M+V_\theta)\,\sigma^2\,, \quad {\sigma = s-1}\,.
\end{align}
Segment~1 runs from the origin to $\gamma(1)=w_M$ with $d\gamma(s)/ds|_{s=1}$ parallel to $T$. Segment~2 bends onto the real axis at $r=L_\text{real}$, $\theta(2)=0$, from where the integration continues along the real axis up to the ultraviolet cutoff.

\subsection{Unequal masses}\label{app:contour-unequal}

With two different pairs of cuts, a single gap midpoint no longer determines the contour, and we construct it numerically. We start from the reference contour of \cref{sec:contour-deformations}: the ray from the origin to $\eta_\text{ref}\sqrt{P^2}$ with $\eta_\text{ref}=\max(\eta,1-\eta)$, a circular arc around the cuts onto the real axis, and the real axis. The deformed contour is written as $w = r\,e^{\textrm{i}\theta(r)}$ with growing radius $r$ and is searched for on the circles $r_k = k\,\Delta r$ with $\Delta r = r_1/K$ and $r_1 = \eta_\text{ref}|\sqrt{P^2}|$, where the integer $K$ sets the radial resolution and is increased adaptively until the narrowest passage between the cuts is resolved. On each circle, the nearest cut crossings on either side of the reference contour define the allowed range
\begin{align}\label{eq:free-range}
	\theta_k^- < \theta(r_k) < \theta_k^+\,.
\end{align}
Any contour that satisfies \labelcref{eq:free-range} on all circles can be moved onto the reference contour without crossing a cut and therefore gives the same integral. On each circle, we place candidate angles $\theta_{k,j}$ inside this range and denote by $d_{k,j}$ the distance of the point $r_k e^{\textrm{i}\theta_{k,j}}$ to the nearest cut. A path through one candidate per circle is assigned the cost
\begin{align}\label{eq:path-cost}
	S &= \sum_k \bar\omega_k\,\sqrt{\Delta r^2 + r_k^2\,(\theta_k-\theta_{k-1})^2}\,, \nonumber\\[2ex]
	\omega(d) &= \max\Bigl(1,\frac{d_c^2}{d^2}\Bigr)\,,
\end{align}
i.e., its length weighted by $\omega$, which grows once the distance to the cuts drops below $d_c$. Here, $\bar\omega_k$ is the average of $\omega(d_{k,j})$ over the candidates on circle $k$ whose angles lie between $\theta_{k-1}$ and $\theta_k$, so that a step is also charged for the candidates it passes on circle $k$, not only for the one it lands on. This avoids contours that come close to a cut between two circles. The minimal cost $V_{k,j}$ of reaching the candidate $(k,j)$ from the origin follows from the recursion
\begin{align}\label{eq:path-dp}
	V_{k,j} = \min_{j'}\Bigl[&V_{k-1,j'}  + \bar\omega_k\,\sqrt{\Delta r^2 + r_k^2\,(\theta_{k,j}-\theta_{k-1,j'})^2}\,\Bigr]\,,
\end{align}
which is evaluated circle by circle. The cheapest path is traced back from the candidate on the real axis at the last circle. The recursion is run twice. The first run determines, in the same sweep, the width of the narrowest passage,
\begin{align}\label{eq:bottleneck}
	B = \max_\text{paths}\,\min_k\, d_k\,.
\end{align}
The second run uses $d_c = 3B$ (restricted to $0.02\,r_1\le d_c\le 0.1\,r_1$) and excludes all candidates with $d_{k,j}<0.85\,B$. The resulting path is straight where there is room and runs through the middle of narrow passages. The optimal path is a sequence of discrete angles $\theta_k$, i.e., a staircase. We smooth it by averaging each $\theta_k$ over the neighbouring circles, within a window that grows with the distance to the cuts, and limit the shift so that the path stays well inside the allowed ranges \labelcref{eq:free-range}. At the smallest radius $r_A$ at which the circular arc $w = r_A e^{\textrm{i}\varphi}$ and the real axis beyond $r_A$ keep a distance of at least $d_c$ to the cuts, the path leaves along this arc onto the real axis and continues there up to the cutoff. Since the quadrature requires a smooth contour, $\theta(r)$ is represented by a cubic Hermite spline, i.e., piecewise cubic polynomials with continuous $\theta$ and $d\theta/dr$, whose knots are the boundaries of contour sections: to increase the performance of the (adaptive) quadrature, we subdivide the path $\theta(r)$, $0\le r\le r_A$, into sections whose length is proportional to the local distance to the cuts, and integrate each section, the arc and the real axis as separate domains.

\section{Numerical Details}\label{app:Numerics}
In this appendix, we give details of the numerical implementation of the quark propagators and the BSEs for both the pseudoscalar and the vector mesons, with a special focus on the decay kernels. All numerical calculations where performed in \textit{Julia} \cite{julia}.

\subsection{Quark propagators}\label{app:Numerics-quark}

On the real Euclidean axis, we solve the gap equation in \cref{fig:quarkgap} with the RL interaction \labelcref{eq:rainbowladder}, renormalised as described in \cref{app:MTmodel}, by fixed-point iteration. 
At complex momenta, we solve the gap equation on the parabolic grid $p^2 = (t+\textrm{i}m)^2$ with $0\le m\le m_\text{max}$. We use $m_\text{max}=0.50$\,GeV for the light and $m_\text{max}=0.70$\,GeV for the strange quark, for which the parabolas do not enclose the first complex-conjugate pole pair of the respective propagator at $\sqrt{p^2} = 0.305\pm0.542\,\textrm{i}$\,GeV and $0.426\pm0.754\,\textrm{i}$\,GeV. 
Routing the external momentum through the quark line keeps the gluon momentum $k$ real, and $(p-k)^2$ then lies inside the parabola of $p^2$, so the iteration closes on the grid. We use the renormalisation constants of the real-axis solution and interpolate the dressing functions bilinearly in $(m,t)$.

\subsection{Bethe-Salpeter equations}\label{app:Numerics-BSE}

The dressing functions of the BSAs depend on $p^2$, $z = \hat p\cdot\hat P$ and $P^2$. 
We work in the rest frame $P=(\sqrt{P^2},0,0,0)$ and write the relative momenta $k\in\{p,q\}$ and the meson loop momentum $l$ of the decay diagrams as
\begin{align}\label{eq:frame}
	k = |k|\bigl(z_k,\; \sqrt{1-z_k^2}\,\hat k_T\bigr)\,,
\end{align}
with $z_k=\hat k\cdot\hat P$ and a real unit vector $\hat k_T$ in the three-dimensional subspace transverse to $P$. On the quadrature nodes, the linear operator $M$ (see \labelcref{eq:mesonBSE-eigenvalue}) becomes a matrix $M_{ij}$ whose rows are labelled by $i=(n,|p|,z_p)$ and columns by $j=(m,|q|,z_q)$, where $n,m$ label the tensor structures. The dimension of the discretised kernel matrix is $N = 8\times 70\times 12 = 6720$ for the vector mesons. We use Gauss-Legendre and Gauss-Chebyshev (second kind) quadratures for the radial and angular nodes, respectively. The column weight $w_j$ contains the radial and Chebyshev weights and the measure of $\int_q$. The direction $\hat p_T$ is fixed, and the result does not depend on this choice. The eigenvalues of the kernel are computed with the Arnoldi method. 

\subsection{Decay kernel factorisation}\label{app:num-factorisation}

In discretised form, the decay kernel \labelcref{eq:Kernelsplit2} reads
\begin{align}\label{eq:Kij-6d}
	M_{ij} = w_j\!\int_{\mathcal C}\!\!d|l|\,|l|^3\!\!\int_{-1}^{1}\!\!dz_l\sqrt{1-z_l^2}\,D_+D_-\,I_{ij}\,,
\end{align}
where $|l|$ runs along the complex contour $\mathcal C$ of \cref{app:contour}, all factors of $2\pi$ are absorbed into the weights, and the angular integral at fixed $\hat p_T$ for $i=(n,|p|,z_p)$ and $j=(m,|q|,z_q)$ is
\begin{align}
	I_{ij} = \int\! d\Omega_l\,d\Omega_q\;L_n^\mu(p,l,P)\,R_m^\mu(q,l,P)\,.
\end{align}
Here, $L_n^\mu$ and $R_m^\mu$ are the left and right halves of the decay diagrams \labelcref{eq:decay-diagrams}, projected onto the tensor structures $n$ and $m$. Amputated, each half is a one-quark exchange between the two meson vertices: $\Gamma_-S(p-l)\Gamma_+$ on the left, see \labelcref{eq:J_l_full}, and $\Gamma_+S(q-l)\Gamma_-$ and $\Gamma_-S(q+l)\Gamma_+$ in the box and crossed box on the right, see \labelcref{eq:J_box}. The two halves are connected only by the loop momentum $l$, which also enters the meson propagators $D_\pm$, and by the contraction of the Lorentz index $\mu$. With $\hat p_T$ fixed, however, both halves depend on $\hat l_T$, so the $\Omega_l$ integral couples them, and each of the $N^2$ matrix elements needs its own six-dimensional integral.

This coupling is removed by fixing $\hat l_T$ instead of $\hat p_T$. At fixed $|p|,z_p,|q|,z_q,|l|,z_l$, the integrand $F_{ij}=L_n^\mu R_m^\mu$ depends on the transverse directions only through their mutual scalar products. It is therefore invariant under a common rotation of $\hat p_T$, $\hat q_T$ and $\hat l_T$, and the integral over any two of them does not depend on the third. Averaging over the third direction gives
\begin{align}
	I_{ij} &= \int\! d\Omega_l\,d\Omega_q\,F_{ij}\Big|_{\hat p_T} = \frac{1}{4\pi}\int\! d\Omega_p\,d\Omega_l\,d\Omega_q\,F_{ij} \nonumber\\[2ex]
	&= \int\! d\Omega_p\,d\Omega_q\,F_{ij}\Big|_{\hat l_T} = \Bigl[\int\! d\Omega_p\,L_n^\mu\Bigr]\Bigl[\int\! d\Omega_q\,R_m^\mu\Bigr]\,,
\end{align}
where the last step uses the fact that, at fixed $\hat l_T$, $L_n$ depends only on $\hat p_T$ and $R_m$ only on $\hat q_T$.

Each pseudoscalar vertex carries one $\gamma^5$. Hence, the Dirac traces in $L_n^\mu$ and $R_m^\mu$ contain no Levi-Civita tensor, and both are linear combinations of the momenta they depend on,
\begin{align}
	L_n^\mu &= \alpha_{P,n}\,P^\mu + \alpha_{p,n}\,p^\mu + \alpha_{l,n}\,l^\mu\,,\nonumber\\
	R_m^\mu &= \beta_{P,m}\,P^\mu + \beta_{q,m}\,q^\mu + \beta_{l,m}\,l^\mu\,,\label{eq:LR-vectors}
\end{align}
where the coefficients $\alpha$ ($\beta$) are scalar functions of the scalar products of $P,p,l$ ($P,q,l$). The Dirac and colour traces are computed with \textit{FormTracer} \cite{Cyrol:2016zqb} and \textit{TensorBases} \cite{Braun:2025gvq}, and the resulting scalar expressions are optimised with \textit{FORM} \cite{Kuipers:2013pba}. 

At this point, the integral for each matrix element factorises into the $(|l|,z_l)$ integral over the product of the two angular integrals over $\Omega_p$ and $\Omega_q$. The factorisation can, however, be exploited even further. After the $\Omega_p$ integration, only $P$ and $l$ remain as fixed vectors. The result is invariant under rotations about $\hat l_T$ and therefore lies in the plane spanned by $P$ and $l_T = l-(l\cdot P)P/P^2$. We take $\hat l_T$ as the polar axis, with $y=\hat p_T\cdot\hat l_T$ and $p_T=p-(p\cdot P)P/P^2$. The azimuthal integral removes the part of $p^\mu$ orthogonal to $P$ and $l_T$, and
\begin{align}
	\int\! d\Omega_p\,L_n^\mu &= A_n\,P^\mu + B_n\,l_T^\mu\,,\nonumber\\
	A_n &= 2\pi\!\int_{-1}^{1}\!\!dy\,\Bigl[\alpha_{P,n} + \frac{(p\cdot P)\,\alpha_{p,n} + (l\cdot P)\,\alpha_{l,n}}{P^2}\Bigr]\,,\nonumber\\
	B_n &= 2\pi\!\int_{-1}^{1}\!\!dy\,\Bigl[\alpha_{l,n} + \frac{p_T\cdot l_T}{l_T^2}\,\alpha_{p,n}\Bigr]\,.\label{eq:AB}
\end{align}
The factor $2\pi$ stems from the two trivially integrated angles. The right half gives $A'_m$ and $B'_m$ in the same way, with $\beta$ and $q$ in place of $\alpha$ and $p$. We denote their values at the nodes $i$ and $j$ by $A_i$, $B_i$ and $A'_j$, $B'_j$. Since $P\cdot l_T=0$, the contraction becomes
\begin{align}
	I_{ij} = P^2\,A_iA'_j + l_T^2\,B_iB'_j\,.
\end{align}
Hence, in each half only the angle $y$ between $p_T$ ($q_T$) and $l_T$ is integrated non-trivially, and a single one-dimensional $y$ integral per half and loop node remains, which gives both $A_n$ and $B_n$. 

We discretise the $(|l|,z_l)$ integral in \labelcref{eq:Kij-6d} with nodes $k$ and weights $\omega_k$, which include the measure and $D_+D_-$. The kernel is then a sum of rank-two terms,
\begin{align}\label{eq:K-rank2}
	M_{ij} = \sum_k \omega_k\bigl[&P^2 A_i(k)\,w_jA'_j(k) + l_{T,k}^2\, B_i(k)\,w_jB'_j(k)\bigr]\,,
\end{align}
i.e., a single product $M=\mathcal A\,\mathcal A'^{T}$ of two $N\times 2N_l$ matrices, where $N_l$ is the number of loop nodes. Filling $\mathcal A$ and $\mathcal A'$ takes $\mathcal O(N N_l N_y)$ evaluations of the integrands of \labelcref{eq:AB}, where $N_y$ is the number of $y$ nodes. This cost grows linearly in $N$, compared to $N^2$ six-dimensional integrals in \labelcref{eq:Kij-6d}. Since $p$ and $q$ live on the same grid, the left and right parts at the same grid point share their quark propagators and vertex functions, which are therefore evaluated only once.

The integrand of the $(|l|,z_l)$ integration is strongly peaked where the contour comes close to the cuts and, on the real segment, for $l\parallel p\parallel q$ with $|l|\approx2|p|$, where the relative momentum of a pion amplitude vanishes. We therefore integrate over $(|l|,z_l)$ adaptively. For \labelcref{eq:K-rank2} to hold, all matrix elements must use the same nodes, so the refinement acts on the matrices $\mathcal A$ and $\mathcal A'$ as a whole. Instead of $z_l$, we integrate over $\theta\in[0,\pi]$ with $z_l=-\cos\theta$, which turns the weight $\sqrt{1-z_l^2}\,dz_l$ with its square-root endpoints into the smooth $\sin^2\theta\,d\theta$. Each contour segment is parametrised by $t\in[0,1]$, and the $(t,\theta)$ domain of each segment is split into rectangular cells. Each cell is integrated with the degree-7 Genz-Malik rule \cite{genz1980remarks}, and its embedded degree-5 rule gives the error estimate. Cells are subdivided until the estimated error of every $(p,q)$ block of the matrix is below $10^{-4}$ of its norm, with an absolute floor for the negligible blocks at large momenta. The $y$ integrals are computed adaptively with Gauss-Kronrod rules for each loop node and grid point.

\subsection{Schlessinger point method}\label{app:spm}
For each meson, $\sqrt{P^2}$ is drawn uniformly from the rectangular windows indicated in \cref{fig:spm-overview}. Through $n$ of the $N$ sampled points $(P_k^2, \lambda_k)$, the SPM constructs the continued fraction $R(P^2)$ with $R(P_k^2) = \lambda_k$~\cite{Schlessinger:1968vsk}, and $R(P^2) = 1$ is solved by Newton iteration. We do this for all $n$-point subsets, or for $2\times10^4$ random ones if there are more than that, and for every $n$ from $N/2$ to $3N/4$ (rounded down). For odd and even $n$, $R$ is a rational function of different degrees ($[k/k]$ and $[k-1/k]$), and the spread of the roots alternates with the parity of $n$. Pooling the roots of all $n$ in this range with equal weight per $n$ includes both types and avoids the choice of a single $n$; leaving out at least a quarter of the points keeps the subsets distinct. The distribution of the roots in $\sqrt{P^2}$ has heavy but continuous tails. We therefore quote the medians of $\operatorname{Re}\sqrt{P^2}$ and $\operatorname{Im}\sqrt{P^2}$ and, as uncertainty, the half-widths of their central 68\% intervals.

\bibliographystyle{utphys_mod}
\bibliography{Resonances_literature}

\end{document}